\documentclass[%
superscriptaddress, %
twocolumn, %
showpacs,
 amsmath,amssymb,
 aps,
prb,
floatfix, longbibliography ]{revtex4-1}
\usepackage{graphicx}
\usepackage{dcolumn}
\usepackage{bm}
\usepackage{soul,color}

\usepackage[colorlinks,linkcolor=blue,anchorcolor=blue,citecolor=blue,urlcolor=blue]{hyperref}
\usepackage{tabularx}
\usepackage{mathtools}

\newcommand{\mb}[1]{\mathbf{#1}}

\begin{document}

\title{Energetic stability and spatial inhomogeneity in the local electronic structure of relaxed twisted trilayer graphene}

\author{Xianqing Lin}
\email[E-mail: ]{xqlin@zjut.edu.cn}
\affiliation{College of Science,
             Zhejiang University of Technology,
             Hangzhou 310023, People's Republic of China}

\author{Cheng Li}
\affiliation{College of Science,
             Zhejiang University of Technology,
             Hangzhou 310023, People's Republic of China}

\author{Kelu Su}
\affiliation{College of Science,
             Zhejiang University of Technology,
             Hangzhou 310023, People's Republic of China}

\author{Jun Ni}
\affiliation{State Key Laboratory of Low-Dimensional Quantum Physics and Frontier Science Center for Quantum Information,
             Department of Physics, Tsinghua University, Beijing 100084,
             People's Republic of China}

\date{\today}

\begin{abstract}
We study the energetic stability and the local electronic structure of the general twisted trilayer graphene (TTG)
with the top and bottom layers rotated with respect to the middle layer respectively by $\theta$ and $\theta'$.
Approximate supercells of the moir\'{e}-of-moir\'{e} superlattices with $\theta$ and $\theta^{\prime}$ within
$1^{\circ}\sim 2^{\circ}$ are established to describe the
structural and electronic properties of relaxed TTG with the periodic boundary condition.
Full relaxation demonstrates that the commensurate TTG with
$\theta=\theta^{\prime}$ has the local minimum total energy ($E_{tol}$) at a fixed $\theta$, while
$E_{tol}$ first reaches a local maximum and begins to drop with decreasing $\theta^{\prime}$ for $\theta^{\prime} < \theta$.
Some regions exhibit enhanced in-plane relaxation in the top and bottom layers but suppressed relaxation in the middle layer and
form a hexagonal network with the moir\'{e}-of-moir\'{e} length scale.
The stacking configurations with the atoms in the three layers vertically aligned at the origin
of the relaxed TTG supercells at $\theta$ around $1.6^{\circ}$ and $\theta^{\prime}$ around $1.4^{\circ}$ have a high
density of states (DOS) near the Fermi level ($E_F$), which can reach that of the mirror symmetric TTG with equal twist angles of about $1.7^{\circ}$.
In contrast, some other stackings can have rather low DOS around $E_F$.
The significant stacking dependence of DOS for some TTG supercells demonstrates that
the local electronic structure of TTG can exhibit
strong spatial inhomogeneity when the twist angles are
slightly away from those of the small supercells with large variations of DOS among different stackings.
Moreover, the structural relaxation of TTG plays a crucial role in the high DOS and its strong stacking dependence.
\end{abstract}

\pacs{%
}



\maketitle


\section{Introduction}

The experimentally precise control of twist angles between consecutive layers in
twisted trilayer graphene (TTG) has introduced superconductivity in
TTG\cite{Tunable-strongly-2021,Electric-field-2021,Pauli-limit-violation-2021}
beyond the magic-angle twisted bilayer graphene (TBG)
\cite{Bistritzer12233,
Cao2018,cao2018unconventional,lu2019superconductors,xie2019spectroscopic,Maximized-electron-2019,Charge-order-2019,uri2020mapping,Strongly-correlated-2020}.
In such superconducting TTG samples,
the relative twist angle ($\theta$) of the top layer and that ($\theta^{\prime}$)
of the bottom layer with respect to the middle layer can share the same value of the magic angle ($\theta_{m}$)
around $1.6^{\circ}$ with the mirror symmetry\cite{Tunable-strongly-2021,Pauli-limit-violation-2021}
or there is a small mismatch of about $0.2^{\circ}$
between $\theta$ and $\theta^{\prime}$ with their values still close to $\theta_{m}$\cite{Electric-field-2021}. The
emergence of superconductivity in TTG implies a high density of states (DOS) around the Fermi level ($E_{F}$),
which has been confirmed by the theoretically calculated
low-energy flat bands in TTG with
$\theta=\theta^{\prime}=\theta_{m}$\cite{Magic-angle-2019,Flatbands-and-2019,Ultraheavy-and-2020,Twisted-symmetric-2021,Band-structure-2021,Stacking-and-2021,Dichotomy-of-2021,
Mirror-symmetry-2021,In-Plane-Critical-2021,Lattice-relaxation-2021,
Emulating-Heavy-2021,Correlated-insulators-2021,Unconventional-superconductivity-2022}.
However, the TTG with mismatched $\theta$ and $\theta^{\prime}$
was predicted to host a rather low DOS at $E_{F}$ when both $\theta$ and
$\theta^{\prime}$ lie in the window of $1^{\circ}\sim 2^{\circ}$\cite{Twisted-Trilayer-2020}.
It is noted that the rigid superlattices of TTG without relaxation
were adopted to obtain such electronic properties\cite{Twisted-Trilayer-2020},
while the structural reconstruction due to the superlattice
relaxation may play a crucial role in enhancing the DOS of TTG with $\theta\neq\theta^{\prime}$, similar to the mirror
symmetric TTG\cite{Ultraheavy-and-2020,Lattice-relaxation-2021} and magic-angle TBG\cite{Dai2016,Nam2017,ShearPhysRevB.98.195432,
Atomicyoo2019atomic,CrucialPhysRevB.99.195419,ContinuumPhysRevB.99.205134,IntrinsicPhysRevB.100.201402,Pressure2020lin}.
Indeed, structural reconstruction has been
observed experimentally in the general TTG with
$\lvert \theta-\theta^{\prime} \rvert$ in the range of
$0.03^{\circ} \sim 0.25^{\circ}$\cite{Twistons-in-2021}. The peculiar
reconstruction patterns were theoretical predicted for TTG and other twisted TMD trilayers\cite{Modeling-mechanical-2020}, while the
energetic stability of TTG due to varying $\theta$ and $\theta^{\prime}$ remains to be revealed.
Therefore, it is crucial to
account for
the structural relaxation to explore systematically the energetic stability and electronic structure of TTG
with general $\theta$ and $\theta^{\prime}$.

In TTG with $\theta\neq\theta^{\prime}$, the moir\'{e}-of-moir\'{e} superlattice with length scale of tens to hundreds
of nanometers can arise with small $\lvert \theta-\theta^{\prime} \rvert$\cite{Modeling-mechanical-2020,Correlated-Insulating-2021}.
Within such a large length scale, the local atomic structure represented by the local stackings between adjacent layers
varies continuously.
The structural relaxation can enhance the spacial variations of the local stacking configurations\cite{Modeling-mechanical-2020}.
Then the local electronic structure, which can be characterized by the local DOS, may also
exhibit strong spatial inhomogeneity, while previous theoretical investigations
of TTG have focused on their global electronic properties\cite{Flatbands-and-2019,Twisted-Trilayer-2020}.
In the trilayer heterostructures of TBG on the hexagonal boron nitride also with two twist angles
\cite{EmergentSharpe605,Intrinsic2020Serlin,Symmetry-breaking-2020,Band-structure-2020,Misalignment2021Lin,Moire2021Shi,Quasiperiodicity2021Mao},
the nonuniform spatial distributions
of the local electronic and topological properties were demonstrated theoretically and can be described by
the supermoir\'{e} picture\cite{Moire2021Shi}.

Here, full relaxation has been done for TTG supercells with general $\theta$ and $\theta^{\prime}$ within
$1^{\circ}\sim 2^{\circ}$ to explore their
energetic and electronic properties. We find that the commensurate TTG with
$\theta=\theta^{\prime}$ has the local minimum total energy ($E_{tol}$) at a fixed $\theta$, while
$E_{tol}$ first reaches a local maximum and begins to drop with decreasing $\theta^{\prime}$ for $\theta^{\prime} < \theta$.
The relaxed TTG with $\theta$ around $1.6^{\circ}$ and $\theta^{\prime}$ around $1.4^{\circ}$ can have large
DOS near $E_{F}$ for some stacking configurations, and the TTG with these twist angles
exhibits strong spatial inhomogeneity in the local electronic structure demonstrated by the stacking dependent
DOS.

The outline of this paper is as follows: In Sec. II we present the structural
configurations of TTG supercells.
The energetic stability and the in-plane structural deformation of fully relaxed TTG are shown in Sec. III.
The stacking dependent electronic structure of relaxed TTG supercells and the spatial distribution of the local electronic structure in
the completely incommensurate TTG are discussed in Sec. IV.
Section V presents the summary and conclusions.

\begin{figure*}[t]
\begin{center}
\includegraphics[width=1.8\columnwidth]{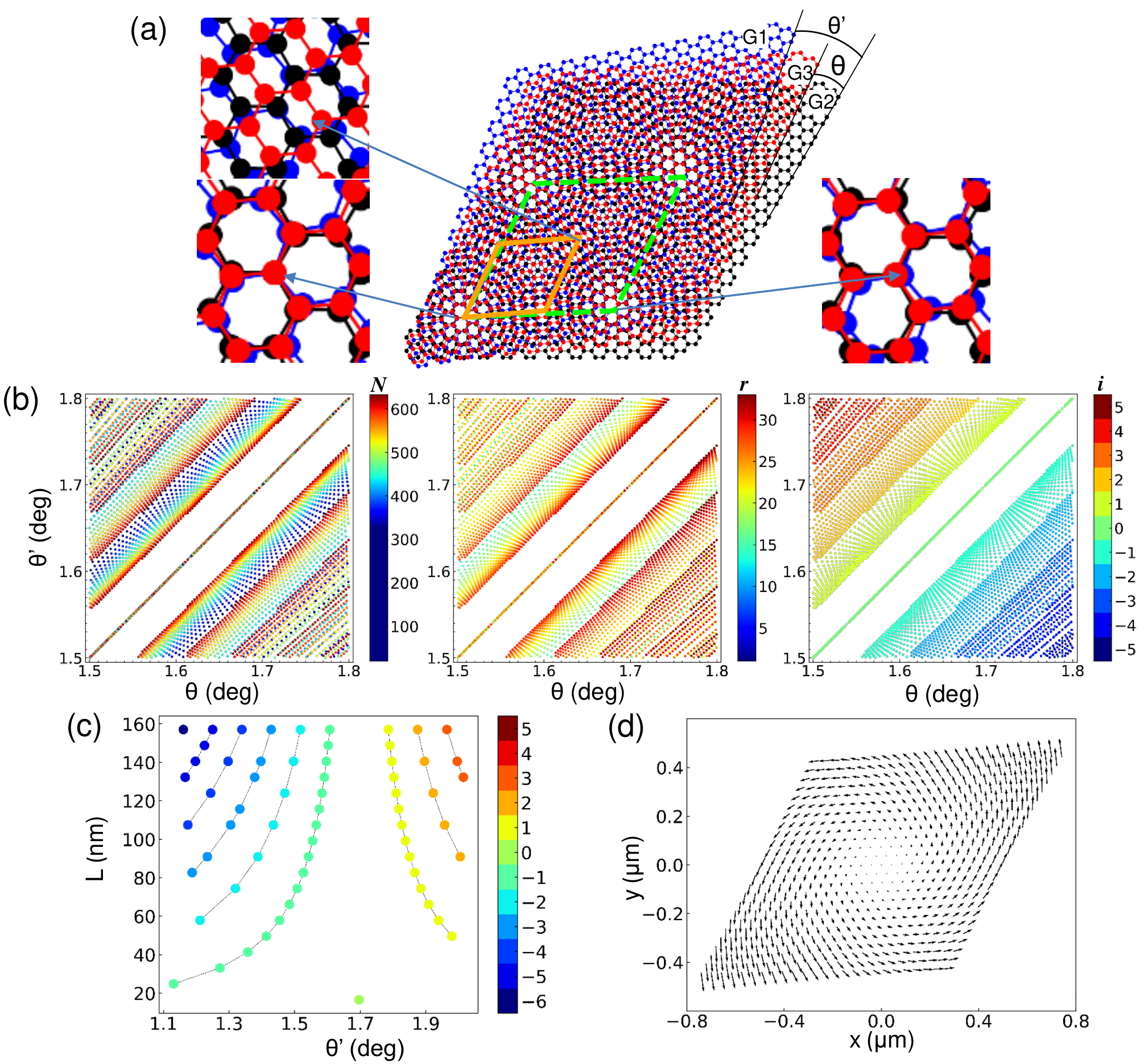}
\end{center}
\caption{(Color online) The geometry of the moir\'{e}-of-moir\'{e} superlattices in TTG. (a)
The schematic view of the TTG with the top (G3) and bottom (G1) layers rotated respectively by $\theta$ and $\theta'$ counterclockwise
with respect to the middle layer (G2).
A moir\'{e} cell between G1 and G2 is represented by the solid lines and that between G2 and G3 is denoted by the dashed lines.
The local stackings at some positions of the superlattices are shown in the insets.
At the origin, the sublattice-A atoms in the three layers are exactly vertically aligned. When G1 is horizontally shifted, the local shift vector between G2 and G1
at the origin is denoted by $\bm{\tau}_{21}$, so $\bm{\tau}_{21} = 0$ for this configuration. The configurations with different $\bm{\tau}_{21}$
can be seen in Fig. S1 of the Supplemental
Material (SM).
At other corners of the moir\'{e} cells, only the sublattice-A atoms in two layers are aligned,
while a strictly periodic supercell consisting of these moir\'{e} cells can be constructed with G1 very slightly strained.
(b) The three integers $N$, $r$, and $i$ determining the supercell geometry.
The supercell size is proportional to $N$, and a supercell consists of
$r \times r$ moir\'{e} cells in G3/G2 and
$(r + i) \times (r + i)$ moir\'{e} cells in G2/G1.
(c) The length ($L$) of the supercell basis vectors as a function of $\theta'$ at $\theta = 1.696^\circ$.
(d) The spatially varying $\bm{\delta'}$ for TTG with $\theta = 1.696^\circ$ and $\theta' = 1.347^\circ$ at
the lattice vectors $\mb{L}$ of the small supercell with $\theta = 1.696^\circ$ and $\theta' = 1.357^\circ$.
\label{fig1}}
\end{figure*}

\section{Supercells of the general TTG}

We first establish the approximate but accurate enough supercells of TTG, which are
used to describe the energetic and electronic properties of the relaxed structures with the periodic boundary condition.
In a TTG, the middle layer (G2) is fixed, and the top (G3) and bottom (G1)
layers are rotated by $\theta$ and $\theta'$ counterclockwise, respectively, as seen in Fig. 1(a).
The unit cell in G2 is spanned by the basis vectors $\mb{a_1} = a(\sqrt{3}/2, -1/2)$ and
$\mb{a_2} = a(\sqrt{3}/2, 1/2)$, where $a = 2.46$ {\AA} is the lattice constant
of graphene.
Double moir\'{e} superlattices arise
between adjacent layers.
Approximate periodic supercells can be built for such double superlattices.
The supercells are taken to be strictly periodic for the superlattice between G3 and G2 (G3/G2) and are spanned by
$\mb{L_1} = N \mb{a_1} + (N + r) \mb{a_2}$ and $\mb{L_2} = T_{60^\circ}\mb{L_1}$ with
$N$ and $r$ positive integers and $T_{60^\circ}$ the rotation
operator by $60^\circ$\cite{LopesdosSantos2007,Commensuration-and-2010,Observation-of-2010,LopesdosSantos2012}.
The twist angle $\theta$ is given by
$\tan (\theta/2) = r/[\sqrt{3} (2n+r)]$
and the length ($L$) of $\mb{L_1}$ can be expressed as $L = ra/[2\sin(\theta/2)]$.
The supercell contains $r \times r$ approximate moir\'{e} cells of G3/G2 as
\begin{equation}
(I - T_{-\theta}) \mb{L_1}= r(\mb{a}_2 - \mb{a}_1).
\end{equation}
We find that $\mb{L_1}$ cannot be an exact lattice vector of G1 when $\theta' \neq \theta$.
However, a very slight biaxial strain ($\epsilon$) in G1 can be introduced to make the supercell also strictly periodic in G1.
Then a position $\mb{r}$ in G2 is transformed to $S\mb{r}$ in G1 with $S = T_{\theta'}/(1 + \epsilon)$.
The supercell is taken to consist of $(r + i) \times (r + i)$ approximate moir\'{e} cells between G2 and G1 (G2/G1) with
\begin{equation}
(S^{-1} - I)\mb{L_1}= (r + i)(\mb{a}_1 - \mb{a}_2),
\end{equation}
where $i$ are small integers.
$\theta'$ and $\epsilon$ can be solved from this equation.
We consider systems with $|\epsilon| < 10^{-4}$, which requires $i$ to be small integers.
Then $\theta'$ is expressed approximately as
\begin{equation}
\sin \theta' \simeq \frac{\sqrt{3} (i + r) (2N + r)}{2 L^2}.
\end{equation}
Zero and positive $i$ give $\theta' \geq \theta$ and a negative $i$ gives $\theta' < \theta$.
The structural parameters of each supercell are thus determined by the three integers $N$, $r$, and $i$.
The possible pairs of $\theta'$ and $\theta$ around 1.65$^\circ$ for supercells with $N \leq 600$ ($L \leq 262$ nm)
are plotted in Fig. 1(b).
The points in Fig. 1(b) form a large number of straight line segments.
Each segment generally has the same $i$, and $N$ and $r$ vary within the segment, with the largest $N$ and $r$
occurring at the end toward small $|\theta' - \theta|$.
There exist some series of supercell configurations with varying $\theta'$ but the same $\theta$.
The possible $L$ oscillates with $\theta'$ at the same $\theta$, as shown in Fig. 1(c) for $\theta = 1.696^\circ$, where
each line has the same $i$. In addition, the reciprocal space of the TTG supercell is shown schematically in Fig. S1 of the SM.

The structural patterns in TTG can be considered as the moir\'{e}
superlattices of the approximate moir\'{e} cells in G3/G2 and those in G2/G1.
A strictly periodic supercell of TTG contains
$r \times r$ moir\'{e} cells in G3/G2 and
$(r + i) \times (r + i)$ moir\'{e} cells in G2/G1. Then
there are $|i| \times |i|$ approximate moir\'{e}-of-moir\'{e} cells within a periodic supercell
for large $r$ and small $i$.

Besides the supercell geometry, the stacking between the double moir\'{e} superlattices
can also influence the energetic and electronic properties of TTG.
The sublattice-A and sublattice-B atoms in a graphene unit cell are located at
$(\mb{a_1} + \mb{a_2})/3$ and $2(\mb{a_1} + \mb{a_2})/3$, respectively.
In a moir\'{e} superlattice, the local stacking between adjacent layers varies continuously and can be characterized by the local in-plane relative shift vectors.
At the origin, the local stacking between G3 and G2 is fixed to be the AA stacking, and the stacking arrangement of a supercell is determined by
the local stacking between G2 and G1 at the origin and thus the corresponding shift vector ($\bm{\tau}_{21}$), as shown in Fig. S2 of the SM.
At an in-plane position $\mb{r}$, the local shift vector between G2 and G1 can be taken as
$\bm{\delta'} = (S^{-1} - I) \mb{r} + \bm{\tau}_{21}$.
In a TTG with twist angles slightly away from those of the small
supercells, the structure becomes completely incommensurate and can be described by the supermoir\'{e} picture\cite{Moire2021Shi}.
In such incommensurate systems, the moir\'{e} superlattices in G3/G2 can still be taken to be periodic with cell vectors of
$\mb{L_1}$ and $\mb{L_2}$, while they are no longer lattice vectors in G1. When the position with AA stacking between G3 and G2 and also between G2 and G1
is chosen as the origin,
the $\bm{\delta'}$ at $\mb{L} = i_1 \mb{L_1} + i_2 \mb{L_2}$ ($i_1$ and $i_2$ are integers) varies slowly with $\mb{L}$.
For example, the supercell with $N = 95$, $r = 5$, and $i = -1$ has $\theta = 1.696^\circ$ and $\theta' = 1.357^\circ$.
For a TTG with the same $\theta$ but $\theta'$ smaller than that of this supercell by just 0.01$^\circ$,
the spatially varying $\bm{\delta'}$ at $\mb{L}$ can be seen in Fig. 1(d).
The $\bm{\delta'}$ can take any vector in a Wigner-Seitz cell of graphene for a TTG sample with dimensions of about 1 $\mu m$.
Then the local electronic properties around the position $\mb{L}$ can be approximately characterized by the supercell with
$\bm{\tau}_{21}$ equal to the $\bm{\delta'}$ at $\mb{L}$.

\section{Structural relaxation of TTG}

\begin{figure}[t]
\begin{center}
\includegraphics[width=1.0\columnwidth]{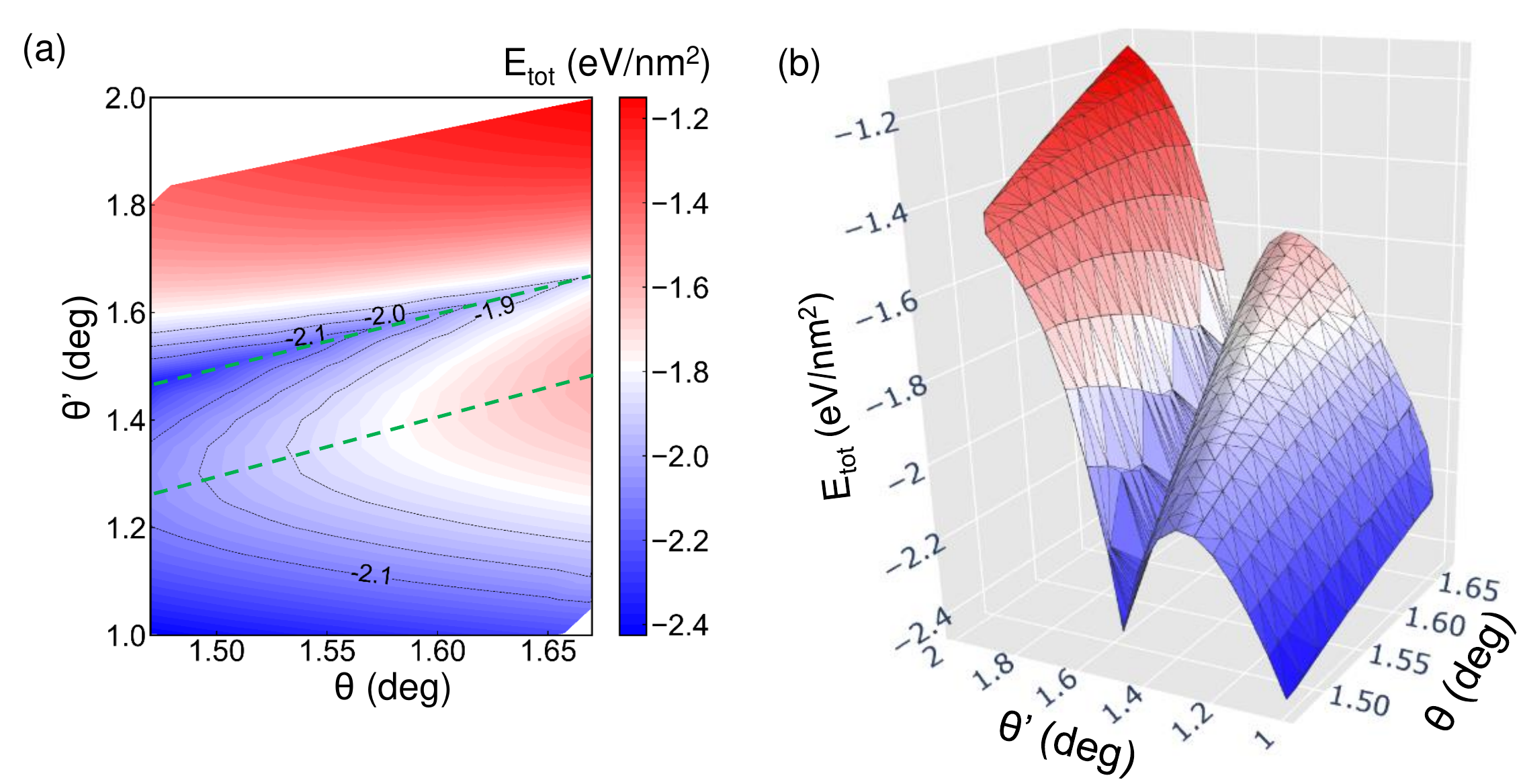}
\end{center}
\caption{(Color online) The total energy ($E_{tot}$) of the relaxed TTG as a function of $\theta$ and $\theta'$.
(a, b) The contour plot and the 3D view of the $E_{tot}$ map.
The green dashed lines in (a) denote systems with $\theta = \theta'$ and
those with the local maximum of $E_{tot}$ at a fixed $\theta$.
\label{fig2}}
\end{figure}

\begin{figure}[t]
\begin{center}
\includegraphics[width=1.0\columnwidth]{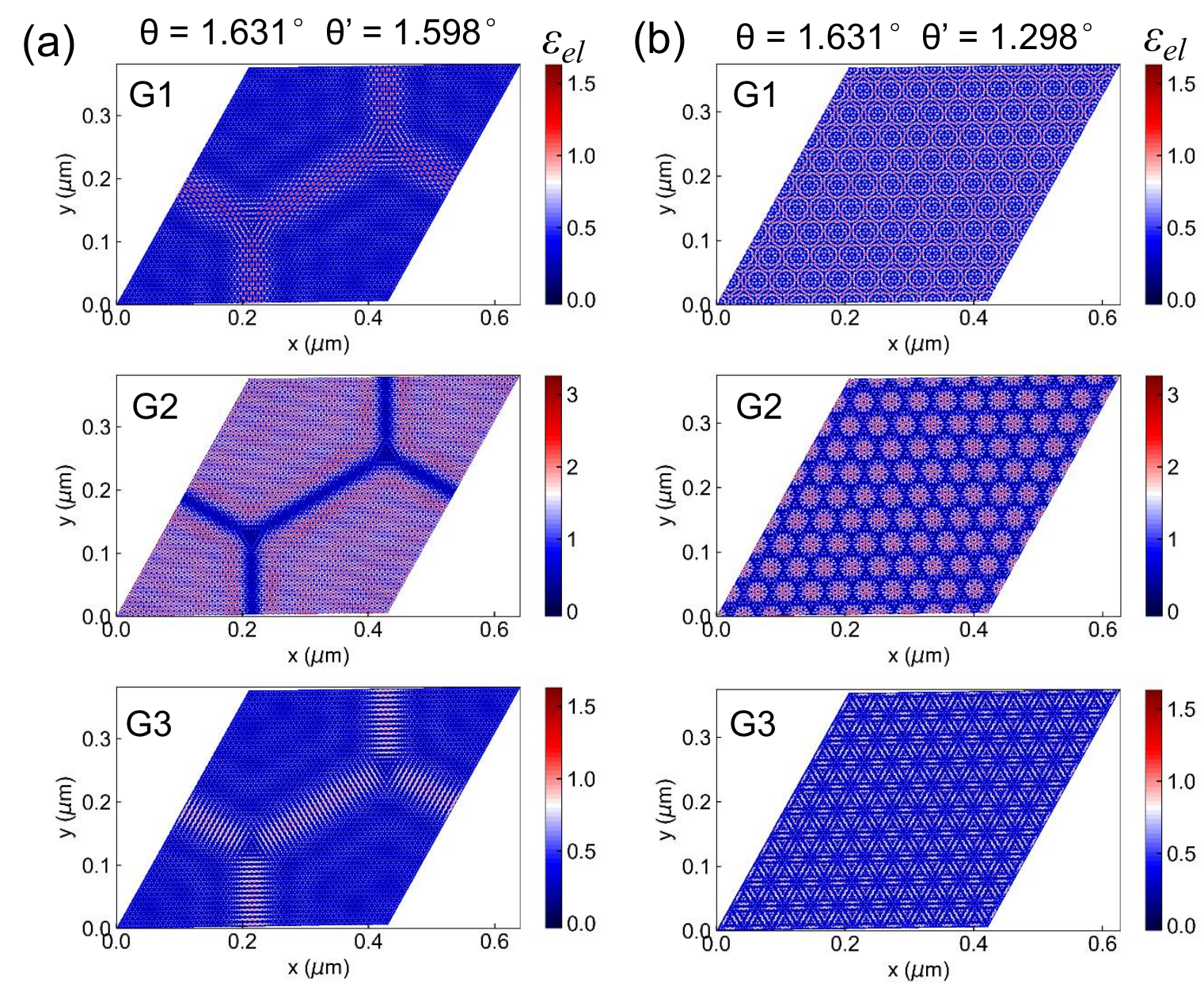}
\end{center}
\caption{(Color online) The spatial distribution of the elastic energy density ($\varepsilon_{el}$) in each layer for a supercell
of TTG with $\theta = 1.631^\circ$ and $\theta' = 1.598^\circ$ ($N = 989$, $r = 50$, $i = -1$) and that with
$\theta = 1.631^\circ$ and $\theta' = 1.298^\circ$ ($N = 969$, $r = 49$, $i = -10$).
$\varepsilon_{el}$ is in the unit of meV per graphene unit cell.
\label{fig3}}
\end{figure}

Since the AB- and BA-like stackings between adjacent graphene layers are
energetically favorable compared with other
stackings\cite{McEuenBLG13,Commensurate2014,Uchida2014,Electronic2014Sep,Spontaneous2014Aug,Effect2015Oct,Origin2015Feb,Wijk2015,Dai2016,Jain2017,Nam2017,Moire2017Aug,Gargiulo2018,
carr2018relaxation,ShearPhysRevB.98.195432,
Twisted-graphene-2018,Atomicyoo2019atomic,CrucialPhysRevB.99.195419,
ContinuumPhysRevB.99.205134,Effective2019lin,Pressure2020lin,
Modeling-mechanical-2020,Tunable-Lattice-2020,Moire-metrology-2021,Localization-of-2021,Emergence-of-2021},
spontaneous in-plane relaxation
occurs in the rigid TTG
due to the energy gain from the
larger domains of energetically favorable local AB- and BA-like stackings.
We have employed the continuum elastic theory to model the in-plane relaxation in the large TTG supercells,
as detailed in the SM.
The displacement field in each layer is expanded in Fourier series to solve the Euler-Lagrange equations, which
minimize the total energy ($E_{tot}$) of a supercell as a functional of the displacement fields.
It is noted that the reciprocal lattice vectors of the supercell with large Fourier components of the displacement fields
are approximately the sum of the small reciprocal lattice vectors of the moir\'{e} cells in G2/G1 and those for G3/G2.

For the fully relaxed TTG, the $E_{tot}$ as a function of $\theta$ and $\theta'$ is displayed in Fig. 2, where
$\theta$ is taken to be around the experimentally realized value of 1.58$^\circ$ and $\theta'$ varies from about 1.0$^\circ$ to 2.0$^\circ$.
We find that the commensurate TTG with $\theta' = \theta$ indeed has the local minimum energy at a fixed $\theta$.
For $\theta' > \theta$, $E_{tot}$ grows fast with increasing $\theta'$. In contrast, $E_{tot}$ first reaches a local maximum and then declines with
decreasing $\theta'$ for $\theta' < \theta$.
So the TTG with $\theta'$ slightly away from $\theta$ may undergo spontaneous structural transformation to reach the commensurate
configuration with equal twist angles, while
the TTG with $\theta'$ rather below $\theta$ can maintain the relative difference between the twist angles.
The appearance of $E_{tot}$ local maximums at $\theta' < \theta$ can be attributed to the competition between
the constructive relaxation in the middle layer G2 at $\theta' = \theta$ and the stronger relaxation in G1 for smaller $\theta'$ with larger
moir\'{e} superlattices in G2/G1.
At  $\theta' = \theta$, the energy favorable AB- or BA-like stackings in G2/G1 and those in G3/G2 are at the same positions, so the
relaxation in G2 can be greatly enhanced due to such constructive interference of the local stackings in the double superlattices.
The constructive relaxation becomes weak for the large $\theta$ with a small commensurate supercell.
Then the $\theta'$ at the $E_{tot}$ local maximum increases with $\theta$.

In the moir\'{e}-of-moir\'{e} superlattices of TTG,
the local stackings exhibit different approximate spatial periods.
In the long period, the in-plane structural relaxation can show strong spatial inhomogeneity,
especially for configurations with small $|i|$ at $\theta'$ close to $\theta$, as shown in Fig. 3.
Most regions in the G2 layer have much larger in-plane strain than that in G1 and G3, while
some continuous positions forming a hexagonal network have greatly suppressed relaxation in G2 but enhanced relaxation in G1 and G3.
Such hexagonal networks are just the $|i| \times |i|$ approximate moir\'{e}-of-moir\'{e} superlattices in a supercell, as clearly demonstrated
for $|i| = 1$ in Fig. 3(a) and for
$|i| = 10$ in Fig. 3(b).
For small $|i|$, the hexagonal networks with large strain in G1 and G3 can be considered as domains walls separating regions with relatively small structural deformation.
Such domains walls may be observed through the flexoelectric effect similar
to that in other twisted graphene layers\cite{Visualization-of-2020,Unraveling-intrinsic-2021}.
In addition, $\bm{\tau}_{21} = \bm{0}$ is adopted to produce the strain maps, so the atoms in the three layers are approximately aligned at the center of each domain.

\begin{figure}[t]
\begin{center}
\includegraphics[width=1.0\columnwidth]{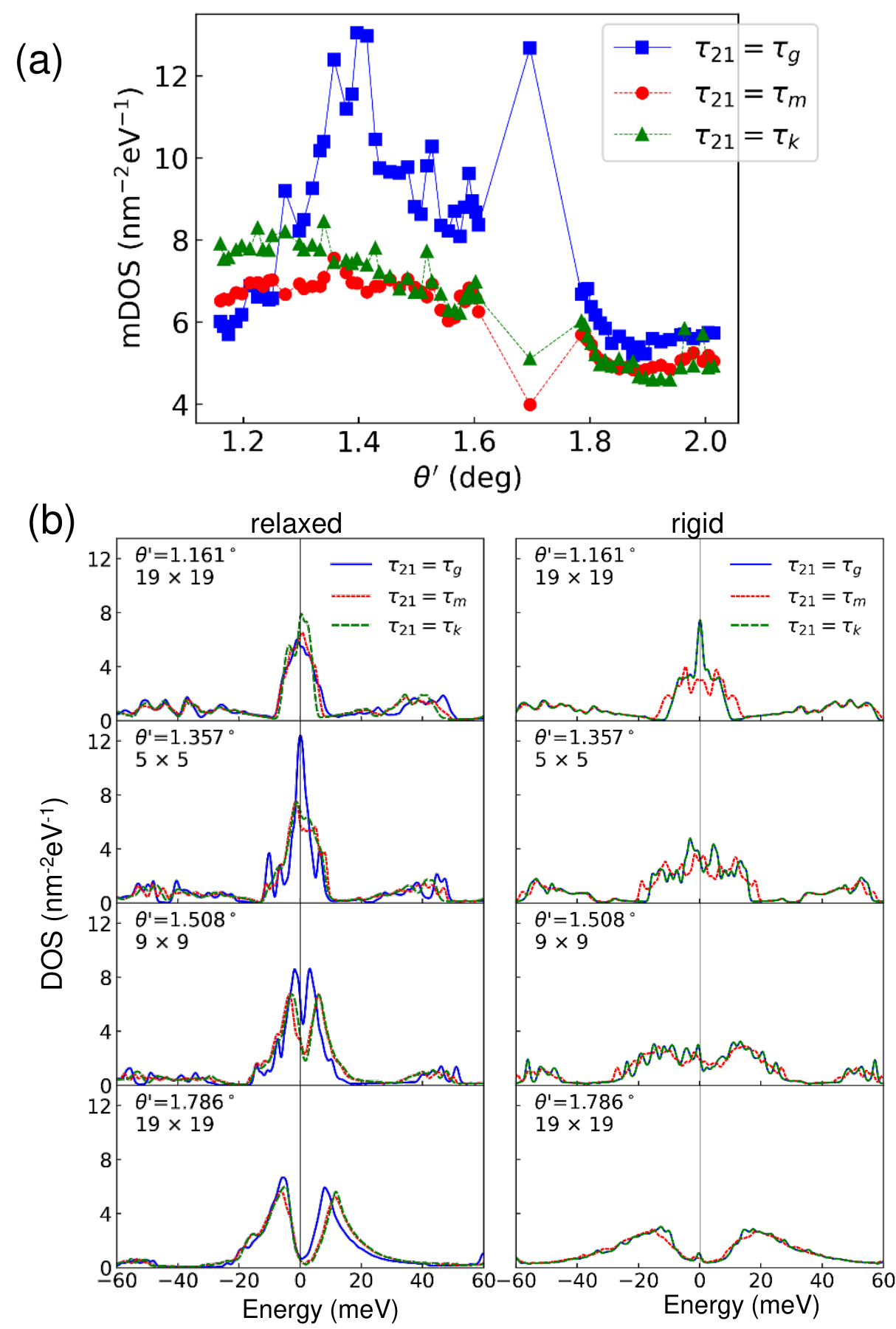}
\end{center}
\caption{(Color online) (a) The mDOS for the three stacking configurations with $\bm{\tau}_{21}$ of
$\bm{\tau}_{g}$, $\bm{\tau}_{k}$, and $\bm{\tau}_{m}$ as a function of $\theta'$
at $\theta = 1.696^\circ$. (b) The DOS for the relaxed and rigid TTG supercells
with different $\theta'$ and stackings at $\theta = 1.696^\circ$.
The size of the supercell containing $r \times r$  moir\'{e} cells of G3/G2 is labeled by $r \times r$.
\label{fig4}}
\end{figure}

\begin{figure}[t]
\begin{center}
\includegraphics[width=1.0\columnwidth]{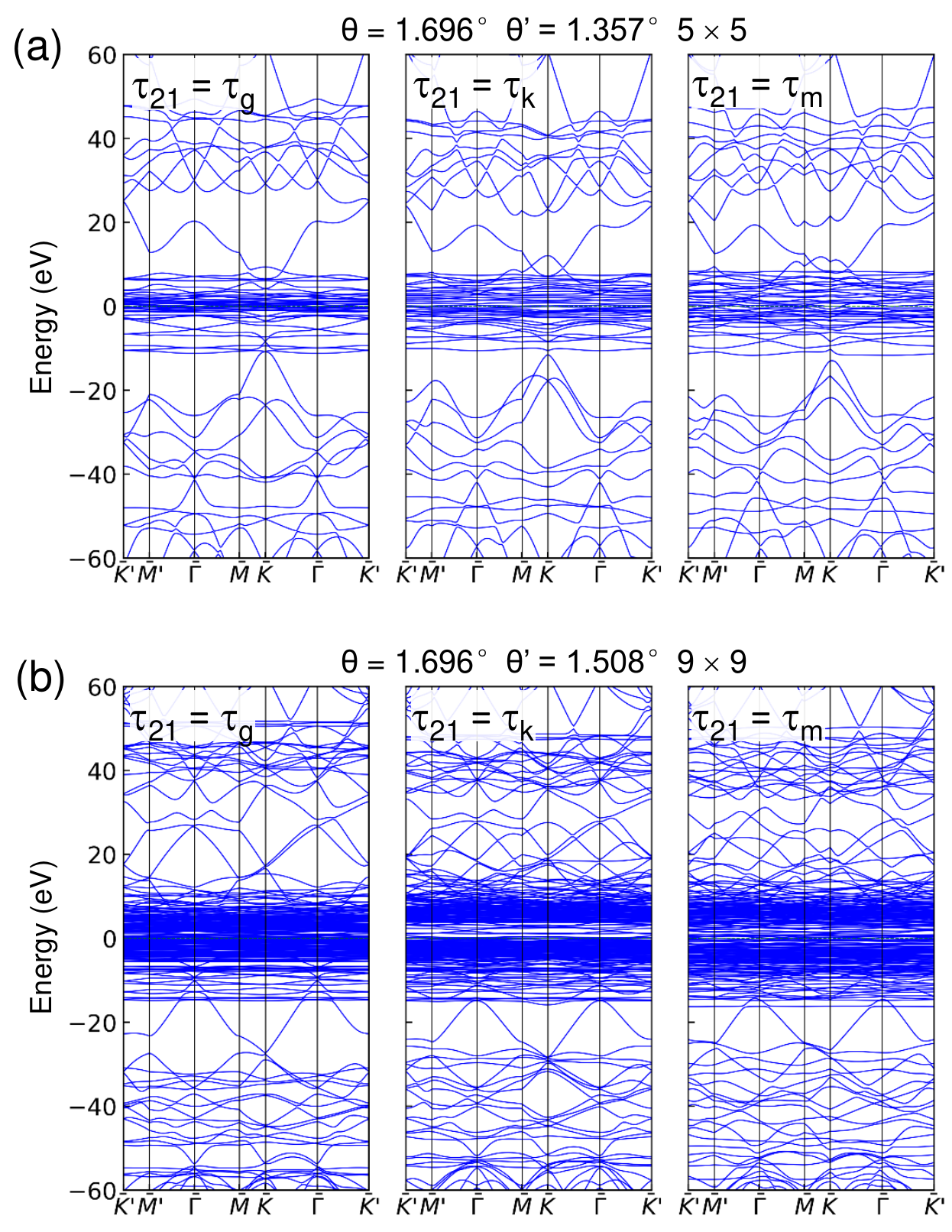}
\end{center}
\caption{(Color online) The band structures of the three stackings of the $5 \times 5$ TTG supercell
at $\theta = 1.696^\circ$ and $\theta' = 1.357^\circ$ (a)
and those of the $9 \times 9$ supercell at $\theta = 1.696^\circ$ and $\theta' = 1.508^\circ$ (b).
\label{fig5}}
\end{figure}

\begin{figure*}[t]
\begin{center}
\includegraphics[width=1.8\columnwidth]{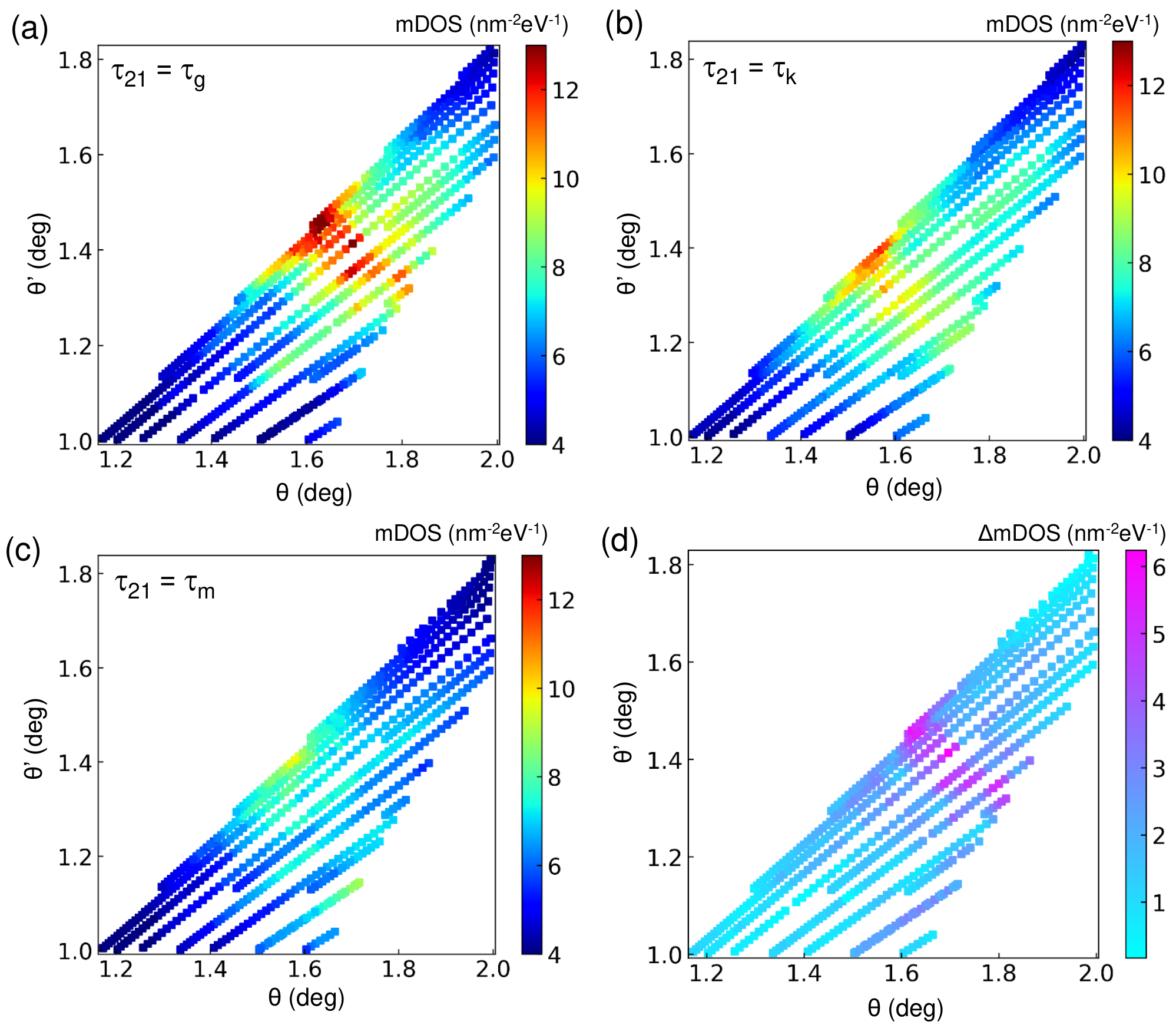}
\end{center}
\caption{(Color online) (a,b,c) The mDOS maps as functions of $\theta$ and $\theta'$ ($\theta > \theta'$) for the three stackings.
The mDOS of the configurations with $L$ smaller than about 90 nm are shown.
(d) The map of the largest difference of mDOS ($\Delta$mDOS) among the three stackings.
\label{fig6}}
\end{figure*}

\begin{figure*}[t]
\begin{center}
\includegraphics[width=1.8\columnwidth]{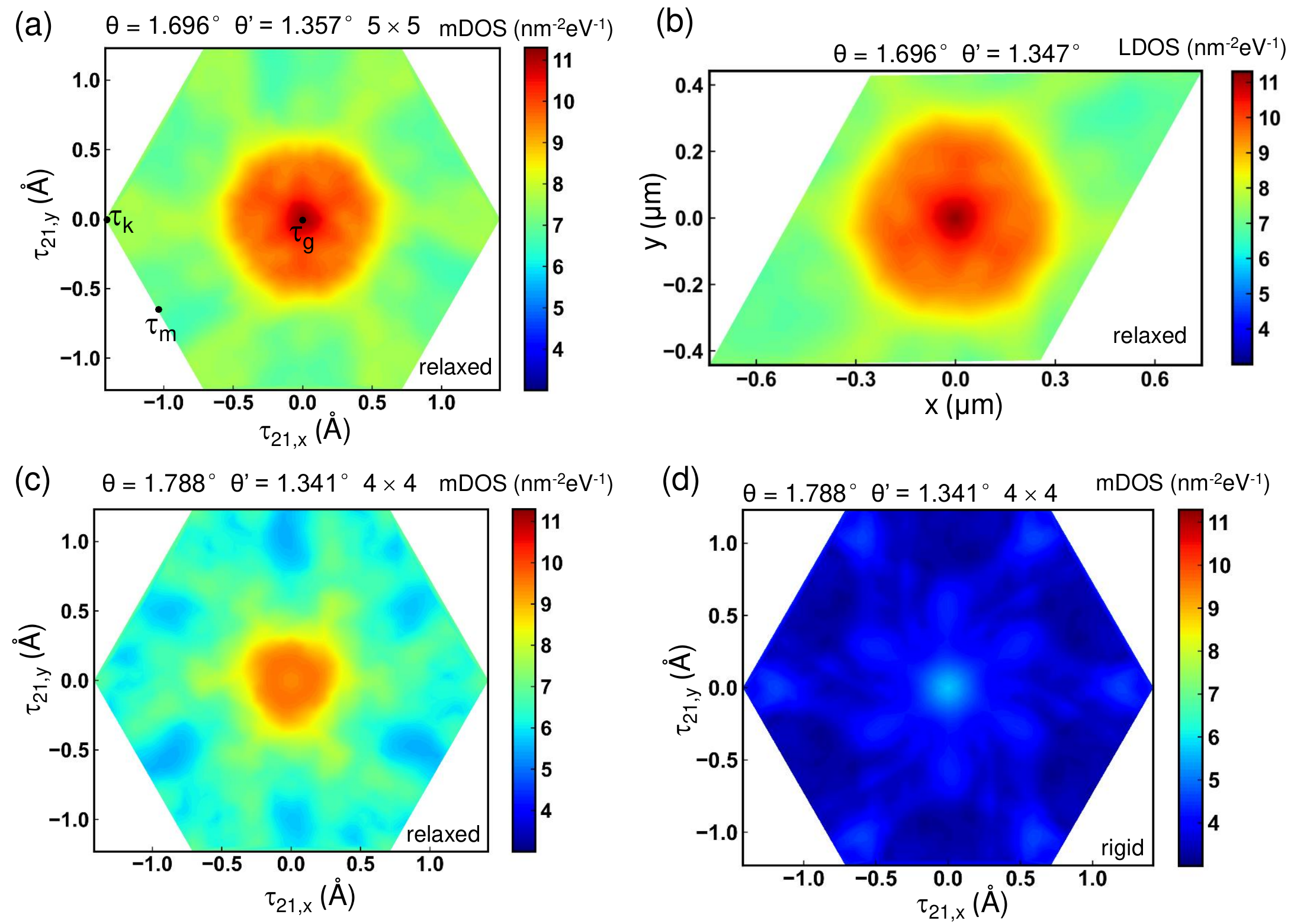}
\end{center}
\caption{(Color online) The mDOS maps as functions of
all the possible stacking $\bm{\tau}_{21}$ in the Wigner-Seitz cell of graphene
for the relaxed $5 \times 5$ supercell at $\theta = 1.696^\circ$ and $\theta' = 1.357^\circ$ (a), and
the relaxed (c) and rigid (d) $4 \times 4$ supercell at $\theta = 1.696^\circ$ and $\theta' = 1.357^\circ$.
The positions of $\bm{\tau}_{g}$, $\bm{\tau}_{k}$, and $\bm{\tau}_{m}$ are labeled in (a).
(b) The spatial map of the local mDOS for the TTG with $\theta = 1.696^\circ$ but $\theta'$ smaller than $1.357^\circ$ by just 0.01$^\circ$.
\label{fig7}}
\end{figure*}

\section{Electronic structure of relaxed TTG}

For a relaxed TTG supercell with given $\theta'$, $\theta$, and $\bm{\tau}_{21}$, the band structure and the density of states can be obtained from the
tight-binding Hamiltonian, which is diagonalized in the plane-wave-like basis functions as detailed in the SM.
In the following, the size of each TTG supercell is labeled by the $r \times r$ moir\'{e} cells of G3/G2 in it.
We first consider configurations with a fixed $\theta = 1.696^\circ$ but varying $\theta'$ and $\bm{\tau}_{21}$.
Figure 4 displays their DOS for the three $\bm{\tau}_{21}$ of $\bm{\tau}_{g} = \bm{0}$, $\bm{\tau}_{m} = -\bm{a}_1/2$,
and $\bm{\tau}_{k} = -(\bm{a}_1 + \bm{a}_2)/3$ and four increasing $\theta'$.
The maximum DOS (mDOS) around $E_F$ for $\bm{\tau}_{g}$ can reach values higher
than 12 $\text{nm}^{-2} \text{eV}^{-1}$ for $\theta'$ around $1.357$ with a single DOS peak at $E_F$.
These mDOS are as high as that of the mirror symmetric TTG at $\theta' = \theta = 1.696^\circ$, as shown in Fig. 4(a).
The mDOS for $\bm{\tau}_{g}$ increases with $\theta'$ from 1.1$^\circ$ to about 1.4$^\circ$, and it tends to become lower for larger $\theta'$ except for
$\theta'$ just equal to $\theta$. The single DOS peak at $E_F$ begins to be split for $\theta' \geq 1.5^\circ$.
In contrast, the mDOSs for $\bm{\tau}_{m}$ and $\bm{\tau}_{k}$ have much smaller values than that for $\bm{\tau}_{g}$ at
$\theta'$ from about 1.3$^\circ$ to $\theta$,
and they just change slightly with $\theta'$.
For $\theta' > \theta$, the mDOSs for the three stacking configurations have similar values, suggesting that the electronic structure of such TTG is approximately
independent of the stacking $\bm{\tau}_{21}$ so that their electronic properties are almost spatially uniform for systems
with twist angles
slightly away from those of the supercells.
We note that the mDOS can be significantly underestimated without the in-plane structural relaxation for all the three stacking configurations, as shown in Fig. 4(b).
At $\theta'$ around 1.36$^\circ$ or larger than that, the mDOS for $\bm{\tau}_{g}$ is just about 4 $\text{nm}^{-2} \text{eV}^{-1}$
without relaxation, which is only one third of that for the relaxed system at
$\theta' = 1.357^\circ$.
This demonstrates that the electronic structure of TTG around $E_F$ is sensitive to the in-plane structural deformation due to the relaxation,
which can enhance the high DOS around $E_F$. In addition, the unrelaxed structures with $\bm{\tau}_{g}$ and
$\bm{\tau}_{k}$ have the same DOS, while the DOS for $\bm{\tau}_{m}$ is different due to the broken $C_{3z}$ symmetry.
For a small $\theta'$ of 1.161$^\circ$, the mDOS without relaxation reaches a rather high value for $\bm{\tau}_{g}$.
Such a high mDOS for unrelaxed structures with a small $\theta'$ around $1.1^\circ$
but a large $\theta$ is consistent with the previous study of the rigid TTG\cite{Twisted-Trilayer-2020}.

The large DOS around $E_F$ in relaxed TTG is also reflected by the band structures of the supercells, as shown in Fig. 5.
For a supercell consisting of a rather large number of moir\'{e} cells in G3/G2 or G2/G1, most subbands around $E_F$ become almost dispersionless.
For the $5 \times 5$ supercell at $\theta' = 1.357^\circ$ and $\theta = 1.696^\circ$, such subbands lie in an energy range of about 20 meV around $E_F$ with the
$\bm{\tau}_{g}$ configuration having flat subbands just at $E_F$ and thus the largest mDOS.
For $\theta' = 1.508^\circ$ and $\theta = 1.696^\circ$ with a larger $9 \times 9$ supercell,
the flat subbands appear in a larger energy range, and there are fewer
subbands at $E_F$ than some other energies close to $E_F$, leading to the split DOS peaks seen in Fig. 4(b).
At energies quite away from $E_F$ by about 40 meV, there are also satellite flat subbands embedded in the dispersive bands, giving rise to the
small DOS peaks away from $E_F$.
In addition, the band structures show that most subbands around $E_F$ connect or cross other bands, while some
bands are separated from nearby bands by direct gaps
smaller than 1 meV, as seen in Fig. 5(a) for the $\bm{\tau}_{k}$
and $\bm{\tau}_{m}$ configurations around the $\bar{K}$ point. Such small direct gaps are present in both relaxed and rigid supercells and are thus caused by
the small approximate supercell.
When the supercell becomes large, the direct gaps tend to vanish, as shown in Fig. 5(b), and
the systems become perfect metal, consistent with the previous calculations of the rigid TTG\cite{Flatbands-and-2019}.

The occurrence of high mDOS and the $\theta'$ with the highest mDOS depend on the value of $\theta$.
Figure 6 illustrates the mDOS maps as functions of $\theta$ and $\theta'$ ($\theta \geq \theta'$)
for supercells with $L \leq 100$ nm for the three
$\bm{\tau}_{21}$.
The configuration with $\theta \leq \theta'$ has the same DOS as that of the corresponding system with  $\theta \geq \theta'$.
For supercells with $\bm{\tau}_{g}$, mDOS can reach 10 $\text{nm}^{-2} \text{eV}^{-1}$ for $\theta$ in the range of
about $1.5^\circ \sim 1.8^\circ$, and the $\theta'$ with the highest mDOS decreases with $\theta$ and lies in the range of about
$1.3^\circ \sim 1.5^\circ$. For $\theta$ and $\theta'$ beyond these ranges, the mDOS is rather low.
The highest mDOS of the supercells with $\bm{\tau}_{k}$ and $\bm{\tau}_{m}$ is smaller than that with $\bm{\tau}_{g}$.
All the mDOS for $\bm{\tau}_{m}$ are lower than 10 $\text{nm}^{-2} \text{eV}^{-1}$,
indicating that the DOS for $\bm{\tau}_{m}$ is weakly related to
$\theta$ and $\theta'$.
For the configurations of $\theta$ and $\theta'$ with high mDOS at $\bm{\tau}_{21} = \bm{\tau}_{g}$,
the differences of mDOS among the three stackings are large, as shown in Fig. 6(d), implying that
the electronic structure of these twist configurations of TTG
can be rather spatially inhomogeneous when
the twist angles are slightly away from those
of the small supercells.

For the $5 \times 5$ supercell at $\theta = 1.696^\circ$ and $\theta' = 1.357^\circ$, we have systematically calculated the variation of mDOS with
all possible stackings, as shown in Fig. 7(a).
The mDOS maintains high values in a rather large region with $|\bm{\tau}_{21}|$ smaller than about 0.5 {\AA}, while it drops fast for large $|\bm{\tau}_{21}|$.
The configurations around $\bm{\tau}_{m}$ have the smallest mDOS.
In a TTG with twist angles slightly away from those of such small commensurate supercells, the structural configurations in different positions can
take various stackings and thus exhibit distinct local electronic structures.
The electronic properties of the incommensurate systems can be described by the supermoir\'{e} picture.
The local DOS at a position $\mb{r}$ is taken approximately as that of the small supercell with the stacking $\bm{\tau}_{21}$
at $\mb{r}$.
Figure 7(b) exhibits the spatial map of the local DOS for the TTG with $\theta = 1.696^\circ$ but $\theta'$ smaller than $1.357^\circ$ by just 0.01$^\circ$.
The spatial variation of the local stacking $\bm{\tau}_{21}$ in this TTG has been shown in Fig. 1(d).
The mDOS has large values in positions $\mb{r}$ around the origin with zero $|\bm{\tau}_{21}|$ for $|\mb{r}|$ smaller than about 0.3 $\mu$m, and
it becomes rather small in other positions, indicating that the local electronic structure is strongly inhomogeneous for this configuration.
Other incommensurate configurations around the small supercells
with large variations of mDOS with $\bm{\tau}_{21}$ [see Fig. 6(d)] can also exhibit inhomogeneous distribution of the local electronic properties,
while the dimension of the region with high mDOS depends on the twist angles.
The $4 \times 4$ supercell at $\theta = 1.788^\circ$ and $\theta' = 1.341^\circ$ has a smaller region of $\bm{\tau}_{21}$ with high mDOS, as shown in
Fig. 7(c). Calculations show that the size of the high-mDOS region is roughly proportional to the mDOS with $\bm{\tau}_{g}$.
It is noted that such spatial inhomogeneity in the local electronic structure can only be observed with the in-plane structural relaxation.
The mDOS of unrelaxed structures remain small for all possible $\bm{\tau}_{21}$, as shown in Fig. 7(d).

\section{Summary and Conclusions}

Supercells of the general TTG with twist angles $\theta$ and $\theta^{\prime}$ within
$1^{\circ}\sim 2^{\circ}$ have been constructed to
perform the full structural relaxation of TTG and obtain the electronic structure of the
relaxed TTG with the periodic boundary condition.
A supercell contains $|i| \times |i|$ approximate moir\'{e}-of-moir\'{e} cells with $i$ a small integer, and its stacking configurations
are characterized by the local in-plane shift vector $\bm{\tau}_{21}$ between G2 and G1 at the origin with the AA stacking between G3 and G2.
The in-plane relaxation of large TTG supercells is performed employing the continuum elastic theory
by solving the Euler-Lagrange equations, where the displacement field in each layer is expanded in Fourier series.
It is shown that the supercell reciprocal lattice vectors with large Fourier components are approximately
the sum of the small reciprocal lattice vectors of the moir\'{e} cells in G2/G1 and those for G3/G2.
We find that the commensurate TTG with
$\theta=\theta^{\prime}$ has the local minimum total energy ($E_{tol}$) at a fixed $\theta$, while
$E_{tol}$ first reaches a local maximum and begins to drop with
decreasing $\theta^{\prime}$ for $\theta' < \theta$.
Some regions exhibit enhanced in-plane relaxation in the top and bottom layers but suppressed relaxation in the middle layer and
form a hexagonal network with the moir\'{e}-of-moir\'{e} length scale.

The band structure and the density of states of a relaxed TTG supercell are obtained by
diagonalizing the Hamiltonian in the plane-wave-like basis functions.
The TTG supercells with the $\bm{\tau}_{g}$ stacking
have high mDOS
at $\theta$ around $1.6^{\circ}$ and $\theta^{\prime}$ around $1.4^{\circ}$, which can reach that of the mirror symmetric TTG with equal twist angles of about $1.7^{\circ}$.
In contrast, the configurations with the $\bm{\tau}_{k}$ and $\bm{\tau}_{m}$ stackings can have rather low mDOS.
Systematic calculations of the variations of mDOS with
all possible stackings demonstrate
the significant stacking dependence of the mDOS for some TTG supercells.
The supermoir\'{e} calculations show that
the local electronic structure of TTG can exhibit
strong spatial inhomogeneity when the twist angles are
slightly away from those of the small supercells with large variations of DOS among different stackings.
Moreover, the structural relaxation of TTG plays a crucial role in the high local DOS and their strong stacking dependence
as the mDOS of unrelaxed structures remain small for all possible stackings.

\label{Acknowledgments}
\begin{acknowledgments}
We gratefully acknowledge valuable discussions with D. Tom\'anek,
H. Xiong, and S. Yin.
This research was supported by
the National Natural Science Foundation of China (Grants No. 11974312 and No. 11774195)
and the Open Research Fund of CNMGE Platform \& NSCC-TJ.
\end{acknowledgments}


\begin{thebibliography}{68}%
\makeatletter
\providecommand \@ifxundefined [1]{%
 \@ifx{#1\undefined}
}%
\providecommand \@ifnum [1]{%
 \ifnum #1\expandafter \@firstoftwo
 \else \expandafter \@secondoftwo
 \fi
}%
\providecommand \@ifx [1]{%
 \ifx #1\expandafter \@firstoftwo
 \else \expandafter \@secondoftwo
 \fi
}%
\providecommand \natexlab [1]{#1}%
\providecommand \enquote  [1]{``#1''}%
\providecommand \bibnamefont  [1]{#1}%
\providecommand \bibfnamefont [1]{#1}%
\providecommand \citenamefont [1]{#1}%
\providecommand \href@noop [0]{\@secondoftwo}%
\providecommand \href [0]{\begingroup \@sanitize@url \@href}%
\providecommand \@href[1]{\@@startlink{#1}\@@href}%
\providecommand \@@href[1]{\endgroup#1\@@endlink}%
\providecommand \@sanitize@url [0]{\catcode `\\12\catcode `\$12\catcode
  `\&12\catcode `\#12\catcode `\^12\catcode `\_12\catcode `\%12\relax}%
\providecommand \@@startlink[1]{}%
\providecommand \@@endlink[0]{}%
\providecommand \url  [0]{\begingroup\@sanitize@url \@url }%
\providecommand \@url [1]{\endgroup\@href {#1}{\urlprefix }}%
\providecommand \urlprefix  [0]{URL }%
\providecommand \Eprint [0]{\href }%
\providecommand \doibase [0]{http://dx.doi.org/}%
\providecommand \selectlanguage [0]{\@gobble}%
\providecommand \bibinfo  [0]{\@secondoftwo}%
\providecommand \bibfield  [0]{\@secondoftwo}%
\providecommand \translation [1]{[#1]}%
\providecommand \BibitemOpen [0]{}%
\providecommand \bibitemStop [0]{}%
\providecommand \bibitemNoStop [0]{.\EOS\space}%
\providecommand \EOS [0]{\spacefactor3000\relax}%
\providecommand \BibitemShut  [1]{\csname bibitem#1\endcsname}%
\let\auto@bib@innerbib\@empty
\bibitem [{\citenamefont {Park}\ \emph {et~al.}(2021)\citenamefont {Park},
  \citenamefont {Cao}, \citenamefont {Watanabe}, \citenamefont {Taniguchi},\
  and\ \citenamefont {Jarillo-Herrero}}]{Tunable-strongly-2021}%
  \BibitemOpen
  \bibfield  {author} {\bibinfo {author} {\bibfnamefont {J.~M.}\ \bibnamefont
  {Park}}, \bibinfo {author} {\bibfnamefont {Y.}~\bibnamefont {Cao}}, \bibinfo
  {author} {\bibfnamefont {K.}~\bibnamefont {Watanabe}}, \bibinfo {author}
  {\bibfnamefont {T.}~\bibnamefont {Taniguchi}}, \ and\ \bibinfo {author}
  {\bibfnamefont {P.}~\bibnamefont {Jarillo-Herrero}},\ }\bibfield  {title}
  {\enquote {\bibinfo {title} {{T}unable strongly coupled superconductivity in
  magic-angle twisted trilayer graphene},}\ }\href {\doibase
  10.1038/s41586-021-03192-0} {\bibfield  {journal} {\bibinfo  {journal}
  {Nature}\ }\textbf {\bibinfo {volume} {590}},\ \bibinfo {pages} {249}
  (\bibinfo {year} {2021})}\BibitemShut {NoStop}%
\bibitem [{\citenamefont {Hao}\ \emph {et~al.}(2021)\citenamefont {Hao},
  \citenamefont {Zimmerman}, \citenamefont {Ledwith}, \citenamefont {Khalaf},
  \citenamefont {Najafabadi}, \citenamefont {Watanabe}, \citenamefont
  {Taniguchi}, \citenamefont {Vishwanath},\ and\ \citenamefont
  {Kim}}]{Electric-field-2021}%
  \BibitemOpen
  \bibfield  {author} {\bibinfo {author} {\bibfnamefont {Z.}~\bibnamefont
  {Hao}}, \bibinfo {author} {\bibfnamefont {A.~M.}\ \bibnamefont {Zimmerman}},
  \bibinfo {author} {\bibfnamefont {P.}~\bibnamefont {Ledwith}}, \bibinfo
  {author} {\bibfnamefont {E.}~\bibnamefont {Khalaf}}, \bibinfo {author}
  {\bibfnamefont {D.~H.}\ \bibnamefont {Najafabadi}}, \bibinfo {author}
  {\bibfnamefont {K.}~\bibnamefont {Watanabe}}, \bibinfo {author}
  {\bibfnamefont {T.}~\bibnamefont {Taniguchi}}, \bibinfo {author}
  {\bibfnamefont {A.}~\bibnamefont {Vishwanath}}, \ and\ \bibinfo {author}
  {\bibfnamefont {P.}~\bibnamefont {Kim}},\ }\bibfield  {title} {\enquote
  {\bibinfo {title} {{E}lectric field–tunable superconductivity in
  alternating-twist magic-angle trilayer graphene},}\ }\href {\doibase
  10.1126/science.abg0399} {\bibfield  {journal} {\bibinfo  {journal}
  {Science}\ }\textbf {\bibinfo {volume} {371}},\ \bibinfo {pages} {1133}
  (\bibinfo {year} {2021})}\BibitemShut {NoStop}%
\bibitem [{\citenamefont {Cao}\ \emph {et~al.}(2021)\citenamefont {Cao},
  \citenamefont {Park}, \citenamefont {Watanabe}, \citenamefont {Taniguchi},\
  and\ \citenamefont {Jarillo-Herrero}}]{Pauli-limit-violation-2021}%
  \BibitemOpen
  \bibfield  {author} {\bibinfo {author} {\bibfnamefont {Y.}~\bibnamefont
  {Cao}}, \bibinfo {author} {\bibfnamefont {J.~M.}\ \bibnamefont {Park}},
  \bibinfo {author} {\bibfnamefont {K.}~\bibnamefont {Watanabe}}, \bibinfo
  {author} {\bibfnamefont {T.}~\bibnamefont {Taniguchi}}, \ and\ \bibinfo
  {author} {\bibfnamefont {P.}~\bibnamefont {Jarillo-Herrero}},\ }\bibfield
  {title} {\enquote {\bibinfo {title} {{P}auli-limit violation and re-entrant
  superconductivity in moir\'{e} graphene},}\ }\href {\doibase
  10.1038/s41586-021-03685-y} {\bibfield  {journal} {\bibinfo  {journal}
  {Nature}\ }\textbf {\bibinfo {volume} {595}},\ \bibinfo {pages} {526}
  (\bibinfo {year} {2021})}\BibitemShut {NoStop}%
\bibitem [{\citenamefont {Bistritzer}\ and\ \citenamefont
  {MacDonald}(2011)}]{Bistritzer12233}%
  \BibitemOpen
  \bibfield  {author} {\bibinfo {author} {\bibfnamefont {R.}~\bibnamefont
  {Bistritzer}}\ and\ \bibinfo {author} {\bibfnamefont {A.~H.}\ \bibnamefont
  {MacDonald}},\ }\bibfield  {title} {\enquote {\bibinfo {title} {{M}oir{\'e}
  bands in twisted double-layer graphene},}\ }\href {\doibase
  10.1073/pnas.1108174108} {\bibfield  {journal} {\bibinfo  {journal} {Proc.
  Natl. Acad. Sci. U.S.A.}\ }\textbf {\bibinfo {volume} {108}},\ \bibinfo
  {pages} {12233} (\bibinfo {year} {2011})}\BibitemShut {NoStop}%
\bibitem [{\citenamefont {Cao}\ \emph {et~al.}(2018{\natexlab{a}})\citenamefont
  {Cao}, \citenamefont {Fatemi}, \citenamefont {Demir}, \citenamefont {Fang},
  \citenamefont {Tomarken}, \citenamefont {Luo}, \citenamefont
  {Sanchez-Yamagishi}, \citenamefont {Watanabe}, \citenamefont {Taniguchi},
  \citenamefont {Kaxiras}, \citenamefont {Ashoori},\ and\ \citenamefont
  {Jarillo-Herrero}}]{Cao2018}%
  \BibitemOpen
  \bibfield  {author} {\bibinfo {author} {\bibfnamefont {Y.}~\bibnamefont
  {Cao}}, \bibinfo {author} {\bibfnamefont {V.}~\bibnamefont {Fatemi}},
  \bibinfo {author} {\bibfnamefont {A.}~\bibnamefont {Demir}}, \bibinfo
  {author} {\bibfnamefont {S.}~\bibnamefont {Fang}}, \bibinfo {author}
  {\bibfnamefont {S.~L.}\ \bibnamefont {Tomarken}}, \bibinfo {author}
  {\bibfnamefont {J.~Y.}\ \bibnamefont {Luo}}, \bibinfo {author} {\bibfnamefont
  {J.~D.}\ \bibnamefont {Sanchez-Yamagishi}}, \bibinfo {author} {\bibfnamefont
  {K.}~\bibnamefont {Watanabe}}, \bibinfo {author} {\bibfnamefont
  {T.}~\bibnamefont {Taniguchi}}, \bibinfo {author} {\bibfnamefont
  {E.}~\bibnamefont {Kaxiras}}, \bibinfo {author} {\bibfnamefont {R.~C.}\
  \bibnamefont {Ashoori}}, \ and\ \bibinfo {author} {\bibfnamefont
  {P.}~\bibnamefont {Jarillo-Herrero}},\ }\bibfield  {title} {\enquote
  {\bibinfo {title} {Correlated insulator behaviour at half-filling in
  magic-angle graphene superlattices},}\ }\href {\doibase 10.1038/nature26154}
  {\bibfield  {journal} {\bibinfo  {journal} {Nature}\ }\textbf {\bibinfo
  {volume} {556}},\ \bibinfo {pages} {80} (\bibinfo {year}
  {2018}{\natexlab{a}})}\BibitemShut {NoStop}%
\bibitem [{\citenamefont {Cao}\ \emph {et~al.}(2018{\natexlab{b}})\citenamefont
  {Cao}, \citenamefont {Fatemi}, \citenamefont {Fang}, \citenamefont
  {Watanabe}, \citenamefont {Taniguchi}, \citenamefont {Kaxiras},\ and\
  \citenamefont {Jarillo-Herrero}}]{cao2018unconventional}%
  \BibitemOpen
  \bibfield  {author} {\bibinfo {author} {\bibfnamefont {Y.}~\bibnamefont
  {Cao}}, \bibinfo {author} {\bibfnamefont {V.}~\bibnamefont {Fatemi}},
  \bibinfo {author} {\bibfnamefont {S.}~\bibnamefont {Fang}}, \bibinfo {author}
  {\bibfnamefont {K.}~\bibnamefont {Watanabe}}, \bibinfo {author}
  {\bibfnamefont {T.}~\bibnamefont {Taniguchi}}, \bibinfo {author}
  {\bibfnamefont {E.}~\bibnamefont {Kaxiras}}, \ and\ \bibinfo {author}
  {\bibfnamefont {P.}~\bibnamefont {Jarillo-Herrero}},\ }\bibfield  {title}
  {\enquote {\bibinfo {title} {Unconventional superconductivity in magic-angle
  graphene superlattices},}\ }\href {http://dx.doi.org/10.1038/nature26160}
  {\bibfield  {journal} {\bibinfo  {journal} {Nature}\ }\textbf {\bibinfo
  {volume} {556}},\ \bibinfo {pages} {43} (\bibinfo {year}
  {2018}{\natexlab{b}})}\BibitemShut {NoStop}%
\bibitem [{\citenamefont {Lu}\ \emph {et~al.}(2019)\citenamefont {Lu},
  \citenamefont {Stepanov}, \citenamefont {Yang}, \citenamefont {Xie},
  \citenamefont {Aamir}, \citenamefont {Das}, \citenamefont {Urgell},
  \citenamefont {Watanabe}, \citenamefont {Taniguchi}, \citenamefont {Zhang},
  \citenamefont {Bachtold}, \citenamefont {MacDonald},\ and\ \citenamefont
  {Efetov}}]{lu2019superconductors}%
  \BibitemOpen
  \bibfield  {author} {\bibinfo {author} {\bibfnamefont {X.}~\bibnamefont
  {Lu}}, \bibinfo {author} {\bibfnamefont {P.}~\bibnamefont {Stepanov}},
  \bibinfo {author} {\bibfnamefont {W.}~\bibnamefont {Yang}}, \bibinfo {author}
  {\bibfnamefont {M.}~\bibnamefont {Xie}}, \bibinfo {author} {\bibfnamefont
  {M.~A.}\ \bibnamefont {Aamir}}, \bibinfo {author} {\bibfnamefont
  {I.}~\bibnamefont {Das}}, \bibinfo {author} {\bibfnamefont {C.}~\bibnamefont
  {Urgell}}, \bibinfo {author} {\bibfnamefont {K.}~\bibnamefont {Watanabe}},
  \bibinfo {author} {\bibfnamefont {T.}~\bibnamefont {Taniguchi}}, \bibinfo
  {author} {\bibfnamefont {G.}~\bibnamefont {Zhang}}, \bibinfo {author}
  {\bibfnamefont {A.}~\bibnamefont {Bachtold}}, \bibinfo {author}
  {\bibfnamefont {A.~H.}\ \bibnamefont {MacDonald}}, \ and\ \bibinfo {author}
  {\bibfnamefont {D.~K.}\ \bibnamefont {Efetov}},\ }\bibfield  {title}
  {\enquote {\bibinfo {title} {Superconductors, orbital magnets and correlated
  states in magic-angle bilayer graphene},}\ }\href {\doibase
  10.1038/s41586-019-1695-0} {\bibfield  {journal} {\bibinfo  {journal}
  {Nature}\ }\textbf {\bibinfo {volume} {574}},\ \bibinfo {pages} {653}
  (\bibinfo {year} {2019})}\BibitemShut {NoStop}%
\bibitem [{\citenamefont {Xie}\ \emph {et~al.}(2019)\citenamefont {Xie},
  \citenamefont {Lian}, \citenamefont {J{\"a}ck}, \citenamefont {Liu},
  \citenamefont {Chiu}, \citenamefont {Watanabe}, \citenamefont {Taniguchi},
  \citenamefont {Bernevig},\ and\ \citenamefont
  {Yazdani}}]{xie2019spectroscopic}%
  \BibitemOpen
  \bibfield  {author} {\bibinfo {author} {\bibfnamefont {Y.}~\bibnamefont
  {Xie}}, \bibinfo {author} {\bibfnamefont {B.}~\bibnamefont {Lian}}, \bibinfo
  {author} {\bibfnamefont {B.}~\bibnamefont {J{\"a}ck}}, \bibinfo {author}
  {\bibfnamefont {X.}~\bibnamefont {Liu}}, \bibinfo {author} {\bibfnamefont
  {C.-L.}\ \bibnamefont {Chiu}}, \bibinfo {author} {\bibfnamefont
  {K.}~\bibnamefont {Watanabe}}, \bibinfo {author} {\bibfnamefont
  {T.}~\bibnamefont {Taniguchi}}, \bibinfo {author} {\bibfnamefont {B.~A.}\
  \bibnamefont {Bernevig}}, \ and\ \bibinfo {author} {\bibfnamefont
  {A.}~\bibnamefont {Yazdani}},\ }\bibfield  {title} {\enquote {\bibinfo
  {title} {Spectroscopic signatures of many-body correlations in magic-angle
  twisted bilayer graphene},}\ }\href {\doibase 10.1038/s41586-019-1422-x}
  {\bibfield  {journal} {\bibinfo  {journal} {Nature}\ }\textbf {\bibinfo
  {volume} {572}},\ \bibinfo {pages} {101} (\bibinfo {year}
  {2019})}\BibitemShut {NoStop}%
\bibitem [{\citenamefont {Kerelsky}\ \emph {et~al.}(2019)\citenamefont
  {Kerelsky}, \citenamefont {McGilly}, \citenamefont {Kennes}, \citenamefont
  {Xian}, \citenamefont {Yankowitz}, \citenamefont {Chen}, \citenamefont
  {Watanabe}, \citenamefont {Taniguchi}, \citenamefont {Hone}, \citenamefont
  {Dean}, \citenamefont {Rubio},\ and\ \citenamefont
  {Pasupathy}}]{Maximized-electron-2019}%
  \BibitemOpen
  \bibfield  {author} {\bibinfo {author} {\bibfnamefont {A.}~\bibnamefont
  {Kerelsky}}, \bibinfo {author} {\bibfnamefont {L.~J.}\ \bibnamefont
  {McGilly}}, \bibinfo {author} {\bibfnamefont {D.~M.}\ \bibnamefont {Kennes}},
  \bibinfo {author} {\bibfnamefont {L.}~\bibnamefont {Xian}}, \bibinfo {author}
  {\bibfnamefont {M.}~\bibnamefont {Yankowitz}}, \bibinfo {author}
  {\bibfnamefont {S.}~\bibnamefont {Chen}}, \bibinfo {author} {\bibfnamefont
  {K.}~\bibnamefont {Watanabe}}, \bibinfo {author} {\bibfnamefont
  {T.}~\bibnamefont {Taniguchi}}, \bibinfo {author} {\bibfnamefont
  {J.}~\bibnamefont {Hone}}, \bibinfo {author} {\bibfnamefont {C.}~\bibnamefont
  {Dean}}, \bibinfo {author} {\bibfnamefont {A.}~\bibnamefont {Rubio}}, \ and\
  \bibinfo {author} {\bibfnamefont {A.~N.}\ \bibnamefont {Pasupathy}},\
  }\bibfield  {title} {\enquote {\bibinfo {title} {Maximized electron
  interactions at the magic angle in twisted bilayer graphene},}\ }\href
  {\doibase 10.1038/s41586-019-1431-9} {\bibfield  {journal} {\bibinfo
  {journal} {Nature}\ }\textbf {\bibinfo {volume} {572}},\ \bibinfo {pages}
  {95} (\bibinfo {year} {2019})}\BibitemShut {NoStop}%
\bibitem [{\citenamefont {Jiang}\ \emph {et~al.}(2019)\citenamefont {Jiang},
  \citenamefont {Lai}, \citenamefont {Watanabe}, \citenamefont {Taniguchi},
  \citenamefont {Haule}, \citenamefont {Mao},\ and\ \citenamefont
  {Andrei}}]{Charge-order-2019}%
  \BibitemOpen
  \bibfield  {author} {\bibinfo {author} {\bibfnamefont {Y.}~\bibnamefont
  {Jiang}}, \bibinfo {author} {\bibfnamefont {X.}~\bibnamefont {Lai}}, \bibinfo
  {author} {\bibfnamefont {K.}~\bibnamefont {Watanabe}}, \bibinfo {author}
  {\bibfnamefont {T.}~\bibnamefont {Taniguchi}}, \bibinfo {author}
  {\bibfnamefont {K.}~\bibnamefont {Haule}}, \bibinfo {author} {\bibfnamefont
  {J.}~\bibnamefont {Mao}}, \ and\ \bibinfo {author} {\bibfnamefont {E.~Y.}\
  \bibnamefont {Andrei}},\ }\bibfield  {title} {\enquote {\bibinfo {title}
  {Charge order and broken rotational symmetry in magic-angle twisted bilayer
  graphene},}\ }\href {\doibase 10.1038/s41586-019-1460-4} {\bibfield
  {journal} {\bibinfo  {journal} {Nature}\ }\textbf {\bibinfo {volume} {573}},\
  \bibinfo {pages} {91} (\bibinfo {year} {2019})}\BibitemShut {NoStop}%
\bibitem [{\citenamefont {Uri}\ \emph {et~al.}(2020)\citenamefont {Uri},
  \citenamefont {Grover}, \citenamefont {Cao}, \citenamefont {Crosse},
  \citenamefont {Bagani}, \citenamefont {Rodan-Legrain}, \citenamefont
  {Myasoedov}, \citenamefont {Watanabe}, \citenamefont {Taniguchi},
  \citenamefont {Moon}, \citenamefont {Koshino}, \citenamefont
  {Jarillo-Herrero},\ and\ \citenamefont {Zeldov}}]{uri2020mapping}%
  \BibitemOpen
  \bibfield  {author} {\bibinfo {author} {\bibfnamefont {A.}~\bibnamefont
  {Uri}}, \bibinfo {author} {\bibfnamefont {S.}~\bibnamefont {Grover}},
  \bibinfo {author} {\bibfnamefont {Y.}~\bibnamefont {Cao}}, \bibinfo {author}
  {\bibfnamefont {J.A.}\ \bibnamefont {Crosse}}, \bibinfo {author}
  {\bibfnamefont {K.}~\bibnamefont {Bagani}}, \bibinfo {author} {\bibfnamefont
  {D.}~\bibnamefont {Rodan-Legrain}}, \bibinfo {author} {\bibfnamefont
  {Y.}~\bibnamefont {Myasoedov}}, \bibinfo {author} {\bibfnamefont
  {K.}~\bibnamefont {Watanabe}}, \bibinfo {author} {\bibfnamefont
  {T.}~\bibnamefont {Taniguchi}}, \bibinfo {author} {\bibfnamefont
  {P.}~\bibnamefont {Moon}}, \bibinfo {author} {\bibfnamefont {M.}~\bibnamefont
  {Koshino}}, \bibinfo {author} {\bibfnamefont {P.}~\bibnamefont
  {Jarillo-Herrero}}, \ and\ \bibinfo {author} {\bibfnamefont {E.}~\bibnamefont
  {Zeldov}},\ }\bibfield  {title} {\enquote {\bibinfo {title} {Mapping the
  twist-angle disorder and {L}andau levels in magic-angle graphene},}\ }\href
  {https://www.nature.com/articles/s41586-020-2255-3} {\bibfield  {journal}
  {\bibinfo  {journal} {Nature}\ }\textbf {\bibinfo {volume} {581}},\ \bibinfo
  {pages} {47} (\bibinfo {year} {2020})}\BibitemShut {NoStop}%
\bibitem [{\citenamefont {Nuckolls}\ \emph {et~al.}(2020)\citenamefont
  {Nuckolls}, \citenamefont {Oh}, \citenamefont {Wong}, \citenamefont {Lian},
  \citenamefont {Watanabe}, \citenamefont {Taniguchi}, \citenamefont
  {Bernevig},\ and\ \citenamefont {Yazdani}}]{Strongly-correlated-2020}%
  \BibitemOpen
  \bibfield  {author} {\bibinfo {author} {\bibfnamefont {K.~P.}\ \bibnamefont
  {Nuckolls}}, \bibinfo {author} {\bibfnamefont {M.}~\bibnamefont {Oh}},
  \bibinfo {author} {\bibfnamefont {D.}~\bibnamefont {Wong}}, \bibinfo {author}
  {\bibfnamefont {B.}~\bibnamefont {Lian}}, \bibinfo {author} {\bibfnamefont
  {K.}~\bibnamefont {Watanabe}}, \bibinfo {author} {\bibfnamefont
  {T.}~\bibnamefont {Taniguchi}}, \bibinfo {author} {\bibfnamefont {B.~A.}\
  \bibnamefont {Bernevig}}, \ and\ \bibinfo {author} {\bibfnamefont
  {A.}~\bibnamefont {Yazdani}},\ }\bibfield  {title} {\enquote {\bibinfo
  {title} {Strongly correlated {C}hern insulators in magic-angle twisted
  bilayer graphene},}\ }\href {\doibase 10.1038/s41586-020-3028-8} {\bibfield
  {journal} {\bibinfo  {journal} {Nature}\ }\textbf {\bibinfo {volume} {588}},\
  \bibinfo {pages} {610} (\bibinfo {year} {2020})}\BibitemShut {NoStop}%
\bibitem [{\citenamefont {Khalaf}\ \emph {et~al.}(2019)\citenamefont {Khalaf},
  \citenamefont {Kruchkov}, \citenamefont {Tarnopolsky},\ and\ \citenamefont
  {Vishwanath}}]{Magic-angle-2019}%
  \BibitemOpen
  \bibfield  {author} {\bibinfo {author} {\bibfnamefont {E.}~\bibnamefont
  {Khalaf}}, \bibinfo {author} {\bibfnamefont {A.~J.}\ \bibnamefont
  {Kruchkov}}, \bibinfo {author} {\bibfnamefont {G.}~\bibnamefont
  {Tarnopolsky}}, \ and\ \bibinfo {author} {\bibfnamefont {A.}~\bibnamefont
  {Vishwanath}},\ }\bibfield  {title} {\enquote {\bibinfo {title} {{M}agic
  angle hierarchy in twisted graphene multilayers},}\ }\href {\doibase
  10.1103/PhysRevB.100.085109} {\bibfield  {journal} {\bibinfo  {journal}
  {Phys. Rev. B}\ }\textbf {\bibinfo {volume} {100}},\ \bibinfo {pages}
  {085109} (\bibinfo {year} {2019})}\BibitemShut {NoStop}%
\bibitem [{\citenamefont {Mora}\ \emph {et~al.}(2019)\citenamefont {Mora},
  \citenamefont {Regnault},\ and\ \citenamefont
  {Bernevig}}]{Flatbands-and-2019}%
  \BibitemOpen
  \bibfield  {author} {\bibinfo {author} {\bibfnamefont {C.}~\bibnamefont
  {Mora}}, \bibinfo {author} {\bibfnamefont {N.}~\bibnamefont {Regnault}}, \
  and\ \bibinfo {author} {\bibfnamefont {B.~A.}\ \bibnamefont {Bernevig}},\
  }\bibfield  {title} {\enquote {\bibinfo {title} {{F}latbands and {P}erfect
  {M}etal in {T}rilayer {M}oir\'e {G}raphene},}\ }\href {\doibase
  10.1103/PhysRevLett.123.026402} {\bibfield  {journal} {\bibinfo  {journal}
  {Phys. Rev. Lett.}\ }\textbf {\bibinfo {volume} {123}},\ \bibinfo {pages}
  {026402} (\bibinfo {year} {2019})}\BibitemShut {NoStop}%
\bibitem [{\citenamefont {Carr}\ \emph {et~al.}(2020)\citenamefont {Carr},
  \citenamefont {Li}, \citenamefont {Zhu}, \citenamefont {Kaxiras},
  \citenamefont {Sachdev},\ and\ \citenamefont
  {Kruchkov}}]{Ultraheavy-and-2020}%
  \BibitemOpen
  \bibfield  {author} {\bibinfo {author} {\bibfnamefont {S.}~\bibnamefont
  {Carr}}, \bibinfo {author} {\bibfnamefont {C.}~\bibnamefont {Li}}, \bibinfo
  {author} {\bibfnamefont {Z.}~\bibnamefont {Zhu}}, \bibinfo {author}
  {\bibfnamefont {E.}~\bibnamefont {Kaxiras}}, \bibinfo {author} {\bibfnamefont
  {S.}~\bibnamefont {Sachdev}}, \ and\ \bibinfo {author} {\bibfnamefont
  {A.}~\bibnamefont {Kruchkov}},\ }\bibfield  {title} {\enquote {\bibinfo
  {title} {{U}ltraheavy and {U}ltrarelativistic {D}irac {Q}uasiparticles in
  {S}andwiched {G}raphenes},}\ }\href {\doibase 10.1021/acs.nanolett.9b04979}
  {\bibfield  {journal} {\bibinfo  {journal} {Nano Lett.}\ }\textbf {\bibinfo
  {volume} {20}},\ \bibinfo {pages} {3030} (\bibinfo {year}
  {2020})}\BibitemShut {NoStop}%
\bibitem [{\citenamefont {C\ifmmode \u{a}\else \u{a}\fi{}lug\ifmmode~\u{a}\else
  \u{a}\fi{}ru}\ \emph {et~al.}(2021)\citenamefont {C\ifmmode \u{a}\else
  \u{a}\fi{}lug\ifmmode~\u{a}\else \u{a}\fi{}ru}, \citenamefont {Xie},
  \citenamefont {Song}, \citenamefont {Lian}, \citenamefont {Regnault},\ and\
  \citenamefont {Bernevig}}]{Twisted-symmetric-2021}%
  \BibitemOpen
  \bibfield  {author} {\bibinfo {author} {\bibfnamefont {D.}~\bibnamefont
  {C\ifmmode \u{a}\else \u{a}\fi{}lug\ifmmode~\u{a}\else \u{a}\fi{}ru}},
  \bibinfo {author} {\bibfnamefont {F.}~\bibnamefont {Xie}}, \bibinfo {author}
  {\bibfnamefont {Z.-D.}\ \bibnamefont {Song}}, \bibinfo {author}
  {\bibfnamefont {B.}~\bibnamefont {Lian}}, \bibinfo {author} {\bibfnamefont
  {N.}~\bibnamefont {Regnault}}, \ and\ \bibinfo {author} {\bibfnamefont
  {B.~A.}\ \bibnamefont {Bernevig}},\ }\bibfield  {title} {\enquote {\bibinfo
  {title} {{T}wisted symmetric trilayer graphene: {S}ingle-particle and
  many-body {H}amiltonians and hidden nonlocal symmetries of trilayer moir\'e
  systems with and without displacement field},}\ }\href {\doibase
  10.1103/PhysRevB.103.195411} {\bibfield  {journal} {\bibinfo  {journal}
  {Phys. Rev. B}\ }\textbf {\bibinfo {volume} {103}},\ \bibinfo {pages}
  {195411} (\bibinfo {year} {2021})}\BibitemShut {NoStop}%
\bibitem [{\citenamefont {Phong}\ \emph {et~al.}(2021)\citenamefont {Phong},
  \citenamefont {Pantale\'on}, \citenamefont {Cea},\ and\ \citenamefont
  {Guinea}}]{Band-structure-2021}%
  \BibitemOpen
  \bibfield  {author} {\bibinfo {author} {\bibfnamefont {V.~T.}\ \bibnamefont
  {Phong}}, \bibinfo {author} {\bibfnamefont {P.~A.}\ \bibnamefont
  {Pantale\'on}}, \bibinfo {author} {\bibfnamefont {T.}~\bibnamefont {Cea}}, \
  and\ \bibinfo {author} {\bibfnamefont {F.}~\bibnamefont {Guinea}},\
  }\bibfield  {title} {\enquote {\bibinfo {title} {{B}and structure and
  superconductivity in twisted trilayer graphene},}\ }\href {\doibase
  10.1103/PhysRevB.104.L121116} {\bibfield  {journal} {\bibinfo  {journal}
  {Phys. Rev. B}\ }\textbf {\bibinfo {volume} {104}},\ \bibinfo {pages}
  {L121116} (\bibinfo {year} {2021})}\BibitemShut {NoStop}%
\bibitem [{\citenamefont {Shin}\ \emph {et~al.}(2021)\citenamefont {Shin},
  \citenamefont {Chittari},\ and\ \citenamefont {Jung}}]{Stacking-and-2021}%
  \BibitemOpen
  \bibfield  {author} {\bibinfo {author} {\bibfnamefont {J.}~\bibnamefont
  {Shin}}, \bibinfo {author} {\bibfnamefont {B.~L.}\ \bibnamefont {Chittari}},
  \ and\ \bibinfo {author} {\bibfnamefont {J.}~\bibnamefont {Jung}},\
  }\bibfield  {title} {\enquote {\bibinfo {title} {{S}tacking and gate-tunable
  topological flat bands, gaps, and anisotropic strip patterns in twisted
  trilayer graphene},}\ }\href {\doibase 10.1103/PhysRevB.104.045413}
  {\bibfield  {journal} {\bibinfo  {journal} {Phys. Rev. B}\ }\textbf {\bibinfo
  {volume} {104}},\ \bibinfo {pages} {045413} (\bibinfo {year}
  {2021})}\BibitemShut {NoStop}%
\bibitem [{\citenamefont {Choi}\ and\ \citenamefont
  {Choi}(2021)}]{Dichotomy-of-2021}%
  \BibitemOpen
  \bibfield  {author} {\bibinfo {author} {\bibfnamefont {Y.~W.}\ \bibnamefont
  {Choi}}\ and\ \bibinfo {author} {\bibfnamefont {H.~J.}\ \bibnamefont
  {Choi}},\ }\bibfield  {title} {\enquote {\bibinfo {title} {{D}ichotomy of
  {E}lectron-{P}honon {C}oupling in {G}raphene {M}oir\'e {F}lat {B}ands},}\
  }\href {\doibase 10.1103/PhysRevLett.127.167001} {\bibfield  {journal}
  {\bibinfo  {journal} {Phys. Rev. Lett.}\ }\textbf {\bibinfo {volume} {127}},\
  \bibinfo {pages} {167001} (\bibinfo {year} {2021})}\BibitemShut {NoStop}%
\bibitem [{\citenamefont {Lei}\ \emph {et~al.}(2021)\citenamefont {Lei},
  \citenamefont {Linhart}, \citenamefont {Qin}, \citenamefont {Libisch},\ and\
  \citenamefont {MacDonald}}]{Mirror-symmetry-2021}%
  \BibitemOpen
  \bibfield  {author} {\bibinfo {author} {\bibfnamefont {C.}~\bibnamefont
  {Lei}}, \bibinfo {author} {\bibfnamefont {L.}~\bibnamefont {Linhart}},
  \bibinfo {author} {\bibfnamefont {W.}~\bibnamefont {Qin}}, \bibinfo {author}
  {\bibfnamefont {F.}~\bibnamefont {Libisch}}, \ and\ \bibinfo {author}
  {\bibfnamefont {A.~H.}\ \bibnamefont {MacDonald}},\ }\bibfield  {title}
  {\enquote {\bibinfo {title} {{M}irror symmetry breaking and lateral stacking
  shifts in twisted trilayer graphene},}\ }\href {\doibase
  10.1103/PhysRevB.104.035139} {\bibfield  {journal} {\bibinfo  {journal}
  {Phys. Rev. B}\ }\textbf {\bibinfo {volume} {104}},\ \bibinfo {pages}
  {035139} (\bibinfo {year} {2021})}\BibitemShut {NoStop}%
\bibitem [{\citenamefont {Qin}\ and\ \citenamefont
  {MacDonald}(2021)}]{In-Plane-Critical-2021}%
  \BibitemOpen
  \bibfield  {author} {\bibinfo {author} {\bibfnamefont {W.}~\bibnamefont
  {Qin}}\ and\ \bibinfo {author} {\bibfnamefont {A.~H.}\ \bibnamefont
  {MacDonald}},\ }\bibfield  {title} {\enquote {\bibinfo {title} {{I}n-{P}lane
  {C}ritical {M}agnetic {F}ields in {M}agic-{A}ngle {T}wisted {T}rilayer
  {G}raphene},}\ }\href {\doibase 10.1103/PhysRevLett.127.097001} {\bibfield
  {journal} {\bibinfo  {journal} {Phys. Rev. Lett.}\ }\textbf {\bibinfo
  {volume} {127}},\ \bibinfo {pages} {097001} (\bibinfo {year}
  {2021})}\BibitemShut {NoStop}%
\bibitem [{\citenamefont {Wu}\ \emph {et~al.}(2021)\citenamefont {Wu},
  \citenamefont {Zhan},\ and\ \citenamefont {Yuan}}]{Lattice-relaxation-2021}%
  \BibitemOpen
  \bibfield  {author} {\bibinfo {author} {\bibfnamefont {Z.}~\bibnamefont
  {Wu}}, \bibinfo {author} {\bibfnamefont {Z.}~\bibnamefont {Zhan}}, \ and\
  \bibinfo {author} {\bibfnamefont {S.}~\bibnamefont {Yuan}},\ }\bibfield
  {title} {\enquote {\bibinfo {title} {{L}attice relaxation, mirror symmetry
  and magnetic field effects on ultraflat bands in twisted trilayer
  graphene},}\ }\href {\doibase 10.1007/s11433-020-1690-4} {\bibfield
  {journal} {\bibinfo  {journal} {Science China Physics, Mechanics \&
  Astronomy}\ }\textbf {\bibinfo {volume} {64}},\ \bibinfo {pages} {267811}
  (\bibinfo {year} {2021})}\BibitemShut {NoStop}%
\bibitem [{\citenamefont {Ramires}\ and\ \citenamefont
  {Lado}(2021)}]{Emulating-Heavy-2021}%
  \BibitemOpen
  \bibfield  {author} {\bibinfo {author} {\bibfnamefont {A.}~\bibnamefont
  {Ramires}}\ and\ \bibinfo {author} {\bibfnamefont {J.~L.}\ \bibnamefont
  {Lado}},\ }\bibfield  {title} {\enquote {\bibinfo {title} {{E}mulating
  {H}eavy {F}ermions in {T}wisted {T}rilayer {G}raphene},}\ }\href {\doibase
  10.1103/PhysRevLett.127.026401} {\bibfield  {journal} {\bibinfo  {journal}
  {Phys. Rev. Lett.}\ }\textbf {\bibinfo {volume} {127}},\ \bibinfo {pages}
  {026401} (\bibinfo {year} {2021})}\BibitemShut {NoStop}%
\bibitem [{\citenamefont {Christos}\ \emph {et~al.}()\citenamefont {Christos},
  \citenamefont {Sachdev},\ and\ \citenamefont
  {S.~Scheurer}}]{Correlated-insulators-2021}%
  \BibitemOpen
  \bibfield  {author} {\bibinfo {author} {\bibfnamefont {M.}~\bibnamefont
  {Christos}}, \bibinfo {author} {\bibfnamefont {S.}~\bibnamefont {Sachdev}}, \
  and\ \bibinfo {author} {\bibfnamefont {M.}~\bibnamefont {S.~Scheurer}},\
  }\bibfield  {title} {\enquote {\bibinfo {title} {{C}orrelated insulators,
  semimetals, and superconductivity in twisted trilayer graphene},}\ }\href
  {https://arxiv.org/abs/2106.02063} {\bibinfo  {journal} {arXiv:2106.02063}\
  }\BibitemShut {NoStop}%
\bibitem [{\citenamefont {Fischer}\ \emph {et~al.}(2022)\citenamefont
  {Fischer}, \citenamefont {Goodwin}, \citenamefont {Mostofi}, \citenamefont
  {Lischner}, \citenamefont {Kennes},\ and\ \citenamefont
  {Klebl}}]{Unconventional-superconductivity-2022}%
  \BibitemOpen
\bibfield  {journal} {  }\bibfield  {author} {\bibinfo {author} {\bibfnamefont
  {A.}~\bibnamefont {Fischer}}, \bibinfo {author} {\bibfnamefont {Z.~A.~H.}\
  \bibnamefont {Goodwin}}, \bibinfo {author} {\bibfnamefont {A.~A.}\
  \bibnamefont {Mostofi}}, \bibinfo {author} {\bibfnamefont {J.}~\bibnamefont
  {Lischner}}, \bibinfo {author} {\bibfnamefont {D.~M.}\ \bibnamefont
  {Kennes}}, \ and\ \bibinfo {author} {\bibfnamefont {L.}~\bibnamefont
  {Klebl}},\ }\bibfield  {title} {\enquote {\bibinfo {title} {{U}nconventional
  superconductivity in magic-angle twisted trilayer graphene},}\ }\href
  {\doibase 10.1038/s41535-021-00410-w} {\bibfield  {journal} {\bibinfo
  {journal} {npj Quantum Materials}\ }\textbf {\bibinfo {volume} {7}},\
  \bibinfo {pages} {5} (\bibinfo {year} {2022})}\BibitemShut {NoStop}%
\bibitem [{\citenamefont {Zhu}\ \emph {et~al.}(2020{\natexlab{a}})\citenamefont
  {Zhu}, \citenamefont {Carr}, \citenamefont {Massatt}, \citenamefont
  {Luskin},\ and\ \citenamefont {Kaxiras}}]{Twisted-Trilayer-2020}%
  \BibitemOpen
  \bibfield  {author} {\bibinfo {author} {\bibfnamefont {Z.}~\bibnamefont
  {Zhu}}, \bibinfo {author} {\bibfnamefont {S.}~\bibnamefont {Carr}}, \bibinfo
  {author} {\bibfnamefont {D.}~\bibnamefont {Massatt}}, \bibinfo {author}
  {\bibfnamefont {M.}~\bibnamefont {Luskin}}, \ and\ \bibinfo {author}
  {\bibfnamefont {E.}~\bibnamefont {Kaxiras}},\ }\bibfield  {title} {\enquote
  {\bibinfo {title} {{T}wisted {T}rilayer {G}raphene: {A} {P}recisely {T}unable
  {P}latform for {C}orrelated {E}lectrons},}\ }\href {\doibase
  10.1103/PhysRevLett.125.116404} {\bibfield  {journal} {\bibinfo  {journal}
  {Phys. Rev. Lett.}\ }\textbf {\bibinfo {volume} {125}},\ \bibinfo {pages}
  {116404} (\bibinfo {year} {2020}{\natexlab{a}})}\BibitemShut {NoStop}%
\bibitem [{\citenamefont {Dai}\ \emph {et~al.}(2016)\citenamefont {Dai},
  \citenamefont {Xiang},\ and\ \citenamefont {Srolovitz}}]{Dai2016}%
  \BibitemOpen
  \bibfield  {author} {\bibinfo {author} {\bibfnamefont {S.}~\bibnamefont
  {Dai}}, \bibinfo {author} {\bibfnamefont {Y.}~\bibnamefont {Xiang}}, \ and\
  \bibinfo {author} {\bibfnamefont {D.~J.}\ \bibnamefont {Srolovitz}},\
  }\bibfield  {title} {\enquote {\bibinfo {title} {Twisted {B}ilayer
  {G}raphene: {M}oir{\'e} with a {T}wist},}\ }\href {\doibase
  10.1021/acs.nanolett.6b02870} {\bibfield  {journal} {\bibinfo  {journal}
  {Nano Lett.}\ }\textbf {\bibinfo {volume} {16}},\ \bibinfo {pages} {5923}
  (\bibinfo {year} {2016})}\BibitemShut {NoStop}%
\bibitem [{\citenamefont {Nam}\ and\ \citenamefont {Koshino}(2017)}]{Nam2017}%
  \BibitemOpen
  \bibfield  {author} {\bibinfo {author} {\bibfnamefont {N.~N.~T.}\
  \bibnamefont {Nam}}\ and\ \bibinfo {author} {\bibfnamefont {M.}~\bibnamefont
  {Koshino}},\ }\bibfield  {title} {\enquote {\bibinfo {title} {Lattice
  relaxation and energy band modulation in twisted bilayer graphene},}\ }\href
  {\doibase 10.1103/PhysRevB.96.075311} {\bibfield  {journal} {\bibinfo
  {journal} {Phys. Rev. B}\ }\textbf {\bibinfo {volume} {96}},\ \bibinfo
  {pages} {075311} (\bibinfo {year} {2017})}\BibitemShut {NoStop}%
\bibitem [{\citenamefont {Lin}\ \emph {et~al.}(2018)\citenamefont {Lin},
  \citenamefont {Liu},\ and\ \citenamefont
  {Tom\'anek}}]{ShearPhysRevB.98.195432}%
  \BibitemOpen
  \bibfield  {author} {\bibinfo {author} {\bibfnamefont {X.}~\bibnamefont
  {Lin}}, \bibinfo {author} {\bibfnamefont {D.}~\bibnamefont {Liu}}, \ and\
  \bibinfo {author} {\bibfnamefont {D.}~\bibnamefont {Tom\'anek}},\ }\bibfield
  {title} {\enquote {\bibinfo {title} {Shear instability in twisted bilayer
  graphene},}\ }\href {\doibase 10.1103/PhysRevB.98.195432} {\bibfield
  {journal} {\bibinfo  {journal} {Phys. Rev. B}\ }\textbf {\bibinfo {volume}
  {98}},\ \bibinfo {pages} {195432} (\bibinfo {year} {2018})}\BibitemShut
  {NoStop}%
\bibitem [{\citenamefont {Yoo}\ \emph {et~al.}(2019)\citenamefont {Yoo},
  \citenamefont {Engelke}, \citenamefont {Carr}, \citenamefont {Fang},
  \citenamefont {Zhang}, \citenamefont {Cazeaux}, \citenamefont {Sung},
  \citenamefont {Hovden}, \citenamefont {Tsen}, \citenamefont {Taniguchi},
  \citenamefont {Watanabe}, \citenamefont {Yi}, \citenamefont {Kim},
  \citenamefont {Luskin}, \citenamefont {Tadmor}, \citenamefont {Kaxiras},\
  and\ \citenamefont {Kim}}]{Atomicyoo2019atomic}%
  \BibitemOpen
  \bibfield  {author} {\bibinfo {author} {\bibfnamefont {H.}~\bibnamefont
  {Yoo}}, \bibinfo {author} {\bibfnamefont {R.}~\bibnamefont {Engelke}},
  \bibinfo {author} {\bibfnamefont {S.}~\bibnamefont {Carr}}, \bibinfo {author}
  {\bibfnamefont {S.}~\bibnamefont {Fang}}, \bibinfo {author} {\bibfnamefont
  {K.}~\bibnamefont {Zhang}}, \bibinfo {author} {\bibfnamefont
  {P.}~\bibnamefont {Cazeaux}}, \bibinfo {author} {\bibfnamefont {S.~H.}\
  \bibnamefont {Sung}}, \bibinfo {author} {\bibfnamefont {R.}~\bibnamefont
  {Hovden}}, \bibinfo {author} {\bibfnamefont {A.~W.}\ \bibnamefont {Tsen}},
  \bibinfo {author} {\bibfnamefont {T.}~\bibnamefont {Taniguchi}}, \bibinfo
  {author} {\bibfnamefont {K.}~\bibnamefont {Watanabe}}, \bibinfo {author}
  {\bibfnamefont {G.-C.}\ \bibnamefont {Yi}}, \bibinfo {author} {\bibfnamefont
  {M.}~\bibnamefont {Kim}}, \bibinfo {author} {\bibfnamefont {M.}~\bibnamefont
  {Luskin}}, \bibinfo {author} {\bibfnamefont {E.~B.}\ \bibnamefont {Tadmor}},
  \bibinfo {author} {\bibfnamefont {E.}~\bibnamefont {Kaxiras}}, \ and\
  \bibinfo {author} {\bibfnamefont {P.}~\bibnamefont {Kim}},\ }\bibfield
  {title} {\enquote {\bibinfo {title} {Atomic and electronic reconstruction at
  the van der {W}aals interface in twisted bilayer graphene},}\ }\href
  {\doibase 10.1038/s41563-019-0346-z} {\bibfield  {journal} {\bibinfo
  {journal} {Nat. Mater.}\ }\textbf {\bibinfo {volume} {18}},\ \bibinfo {pages}
  {448} (\bibinfo {year} {2019})}\BibitemShut {NoStop}%
\bibitem [{\citenamefont {Lucignano}\ \emph {et~al.}(2019)\citenamefont
  {Lucignano}, \citenamefont {Alf\`e}, \citenamefont {Cataudella},
  \citenamefont {Ninno},\ and\ \citenamefont
  {Cantele}}]{CrucialPhysRevB.99.195419}%
  \BibitemOpen
  \bibfield  {author} {\bibinfo {author} {\bibfnamefont {P.}~\bibnamefont
  {Lucignano}}, \bibinfo {author} {\bibfnamefont {D.}~\bibnamefont {Alf\`e}},
  \bibinfo {author} {\bibfnamefont {V.}~\bibnamefont {Cataudella}}, \bibinfo
  {author} {\bibfnamefont {D.}~\bibnamefont {Ninno}}, \ and\ \bibinfo {author}
  {\bibfnamefont {G.}~\bibnamefont {Cantele}},\ }\bibfield  {title} {\enquote
  {\bibinfo {title} {Crucial role of atomic corrugation on the flat bands and
  energy gaps of twisted bilayer graphene at the magic angle
  $\ensuremath{\theta}\ensuremath{\sim}1.{08}^{\ensuremath{\circ}}$},}\ }\href
  {\doibase 10.1103/PhysRevB.99.195419} {\bibfield  {journal} {\bibinfo
  {journal} {Phys. Rev. B}\ }\textbf {\bibinfo {volume} {99}},\ \bibinfo
  {pages} {195419} (\bibinfo {year} {2019})}\BibitemShut {NoStop}%
\bibitem [{\citenamefont {Guinea}\ and\ \citenamefont
  {Walet}(2019)}]{ContinuumPhysRevB.99.205134}%
  \BibitemOpen
  \bibfield  {author} {\bibinfo {author} {\bibfnamefont {F.}~\bibnamefont
  {Guinea}}\ and\ \bibinfo {author} {\bibfnamefont {N.~R.}\ \bibnamefont
  {Walet}},\ }\bibfield  {title} {\enquote {\bibinfo {title} {Continuum models
  for twisted bilayer graphene: {E}ffect of lattice deformation and hopping
  parameters},}\ }\href {\doibase 10.1103/PhysRevB.99.205134} {\bibfield
  {journal} {\bibinfo  {journal} {Phys. Rev. B}\ }\textbf {\bibinfo {volume}
  {99}},\ \bibinfo {pages} {205134} (\bibinfo {year} {2019})}\BibitemShut
  {NoStop}%
\bibitem [{\citenamefont {Choi}\ and\ \citenamefont
  {Choi}(2019)}]{IntrinsicPhysRevB.100.201402}%
  \BibitemOpen
  \bibfield  {author} {\bibinfo {author} {\bibfnamefont {Y.~W.}\ \bibnamefont
  {Choi}}\ and\ \bibinfo {author} {\bibfnamefont {H.~J.}\ \bibnamefont
  {Choi}},\ }\bibfield  {title} {\enquote {\bibinfo {title} {Intrinsic band gap
  and electrically tunable flat bands in twisted double bilayer graphene},}\
  }\href {\doibase 10.1103/PhysRevB.100.201402} {\bibfield  {journal} {\bibinfo
   {journal} {Phys. Rev. B}\ }\textbf {\bibinfo {volume} {100}},\ \bibinfo
  {pages} {201402} (\bibinfo {year} {2019})}\BibitemShut {NoStop}%
\bibitem [{\citenamefont {Lin}\ \emph {et~al.}(2020)\citenamefont {Lin},
  \citenamefont {Zhu},\ and\ \citenamefont {Ni}}]{Pressure2020lin}%
  \BibitemOpen
  \bibfield  {author} {\bibinfo {author} {\bibfnamefont {X.}~\bibnamefont
  {Lin}}, \bibinfo {author} {\bibfnamefont {H.}~\bibnamefont {Zhu}}, \ and\
  \bibinfo {author} {\bibfnamefont {J.}~\bibnamefont {Ni}},\ }\bibfield
  {title} {\enquote {\bibinfo {title} {Pressure-induced gap modulation and
  topological transitions in twisted bilayer and twisted double bilayer
  graphene},}\ }\href {\doibase 10.1103/PhysRevB.101.155405} {\bibfield
  {journal} {\bibinfo  {journal} {Phys. Rev. B}\ }\textbf {\bibinfo {volume}
  {101}},\ \bibinfo {pages} {155405} (\bibinfo {year} {2020})}\BibitemShut
  {NoStop}%
\bibitem [{\citenamefont {Turkel}\ \emph {et~al.}()\citenamefont {Turkel},
  \citenamefont {Swann}, \citenamefont {Zhu}, \citenamefont {Christos},
  \citenamefont {Watanabe}, \citenamefont {Taniguchi}, \citenamefont {Sachdev},
  \citenamefont {S.~Scheurer}, \citenamefont {Kaxiras}, \citenamefont
  {R.~Dean},\ and\ \citenamefont {N.~Pasupathy}}]{Twistons-in-2021}%
  \BibitemOpen
  \bibfield  {author} {\bibinfo {author} {\bibfnamefont {S.}~\bibnamefont
  {Turkel}}, \bibinfo {author} {\bibfnamefont {J.}~\bibnamefont {Swann}},
  \bibinfo {author} {\bibfnamefont {Z.}~\bibnamefont {Zhu}}, \bibinfo {author}
  {\bibfnamefont {M.}~\bibnamefont {Christos}}, \bibinfo {author}
  {\bibfnamefont {K.}~\bibnamefont {Watanabe}}, \bibinfo {author}
  {\bibfnamefont {T.}~\bibnamefont {Taniguchi}}, \bibinfo {author}
  {\bibfnamefont {S.}~\bibnamefont {Sachdev}}, \bibinfo {author} {\bibfnamefont
  {M.}~\bibnamefont {S.~Scheurer}}, \bibinfo {author} {\bibfnamefont
  {E.}~\bibnamefont {Kaxiras}}, \bibinfo {author} {\bibfnamefont
  {C.}~\bibnamefont {R.~Dean}}, \ and\ \bibinfo {author} {\bibfnamefont
  {A.}~\bibnamefont {N.~Pasupathy}},\ }\bibfield  {title} {\enquote {\bibinfo
  {title} {{T}wistons in a {S}ea of {M}agic},}\ }\href
  {https://arxiv.org/abs/2109.12631} {\bibinfo  {journal} {arXiv:2109.12631}\
  }\BibitemShut {NoStop}%
\bibitem [{\citenamefont {Zhu}\ \emph {et~al.}(2020{\natexlab{b}})\citenamefont
  {Zhu}, \citenamefont {Cazeaux}, \citenamefont {Luskin},\ and\ \citenamefont
  {Kaxiras}}]{Modeling-mechanical-2020}%
  \BibitemOpen
\bibfield  {journal} {  }\bibfield  {author} {\bibinfo {author} {\bibfnamefont
  {Z.}~\bibnamefont {Zhu}}, \bibinfo {author} {\bibfnamefont {P.}~\bibnamefont
  {Cazeaux}}, \bibinfo {author} {\bibfnamefont {M.}~\bibnamefont {Luskin}}, \
  and\ \bibinfo {author} {\bibfnamefont {E.}~\bibnamefont {Kaxiras}},\
  }\bibfield  {title} {\enquote {\bibinfo {title} {{M}odeling mechanical
  relaxation in incommensurate trilayer van der {W}aals heterostructures},}\
  }\href {\doibase 10.1103/PhysRevB.101.224107} {\bibfield  {journal} {\bibinfo
   {journal} {Phys. Rev. B}\ }\textbf {\bibinfo {volume} {101}},\ \bibinfo
  {pages} {224107} (\bibinfo {year} {2020}{\natexlab{b}})}\BibitemShut
  {NoStop}%
\bibitem [{\citenamefont {Zhang}\ \emph {et~al.}(2021)\citenamefont {Zhang},
  \citenamefont {Tsai}, \citenamefont {Zhu}, \citenamefont {Ren}, \citenamefont
  {Luo}, \citenamefont {Carr}, \citenamefont {Luskin}, \citenamefont
  {Kaxiras},\ and\ \citenamefont {Wang}}]{Correlated-Insulating-2021}%
  \BibitemOpen
  \bibfield  {author} {\bibinfo {author} {\bibfnamefont {X.}~\bibnamefont
  {Zhang}}, \bibinfo {author} {\bibfnamefont {K.-T.}\ \bibnamefont {Tsai}},
  \bibinfo {author} {\bibfnamefont {Z.}~\bibnamefont {Zhu}}, \bibinfo {author}
  {\bibfnamefont {W.}~\bibnamefont {Ren}}, \bibinfo {author} {\bibfnamefont
  {Y.}~\bibnamefont {Luo}}, \bibinfo {author} {\bibfnamefont {S.}~\bibnamefont
  {Carr}}, \bibinfo {author} {\bibfnamefont {M.}~\bibnamefont {Luskin}},
  \bibinfo {author} {\bibfnamefont {E.}~\bibnamefont {Kaxiras}}, \ and\
  \bibinfo {author} {\bibfnamefont {K.}~\bibnamefont {Wang}},\ }\bibfield
  {title} {\enquote {\bibinfo {title} {{C}orrelated {I}nsulating {S}tates and
  {T}ransport {S}ignature of {S}uperconductivity in {T}wisted {T}rilayer
  {G}raphene {S}uperlattices},}\ }\href {\doibase
  10.1103/PhysRevLett.127.166802} {\bibfield  {journal} {\bibinfo  {journal}
  {Phys. Rev. Lett.}\ }\textbf {\bibinfo {volume} {127}},\ \bibinfo {pages}
  {166802} (\bibinfo {year} {2021})}\BibitemShut {NoStop}%
\bibitem [{\citenamefont {Sharpe}\ \emph {et~al.}(2019)\citenamefont {Sharpe},
  \citenamefont {Fox}, \citenamefont {Barnard}, \citenamefont {Finney},
  \citenamefont {Watanabe}, \citenamefont {Taniguchi}, \citenamefont
  {Kastner},\ and\ \citenamefont {Goldhaber-Gordon}}]{EmergentSharpe605}%
  \BibitemOpen
  \bibfield  {author} {\bibinfo {author} {\bibfnamefont {A.~L.}\ \bibnamefont
  {Sharpe}}, \bibinfo {author} {\bibfnamefont {E.~J.}\ \bibnamefont {Fox}},
  \bibinfo {author} {\bibfnamefont {A.~W.}\ \bibnamefont {Barnard}}, \bibinfo
  {author} {\bibfnamefont {J.}~\bibnamefont {Finney}}, \bibinfo {author}
  {\bibfnamefont {K.}~\bibnamefont {Watanabe}}, \bibinfo {author}
  {\bibfnamefont {T.}~\bibnamefont {Taniguchi}}, \bibinfo {author}
  {\bibfnamefont {M.~A.}\ \bibnamefont {Kastner}}, \ and\ \bibinfo {author}
  {\bibfnamefont {D.}~\bibnamefont {Goldhaber-Gordon}},\ }\bibfield  {title}
  {\enquote {\bibinfo {title} {Emergent ferromagnetism near three-quarters
  filling in twisted bilayer graphene},}\ }\href {\doibase
  10.1126/science.aaw3780} {\bibfield  {journal} {\bibinfo  {journal}
  {Science}\ }\textbf {\bibinfo {volume} {365}},\ \bibinfo {pages} {605}
  (\bibinfo {year} {2019})}\BibitemShut {NoStop}%
\bibitem [{\citenamefont {Serlin}\ \emph {et~al.}(2020)\citenamefont {Serlin},
  \citenamefont {Tschirhart}, \citenamefont {Polshyn}, \citenamefont {Zhang},
  \citenamefont {Zhu}, \citenamefont {Watanabe}, \citenamefont {Taniguchi},
  \citenamefont {Balents},\ and\ \citenamefont {Young}}]{Intrinsic2020Serlin}%
  \BibitemOpen
  \bibfield  {author} {\bibinfo {author} {\bibfnamefont {M.}~\bibnamefont
  {Serlin}}, \bibinfo {author} {\bibfnamefont {C.~L.}\ \bibnamefont
  {Tschirhart}}, \bibinfo {author} {\bibfnamefont {H.}~\bibnamefont {Polshyn}},
  \bibinfo {author} {\bibfnamefont {Y.}~\bibnamefont {Zhang}}, \bibinfo
  {author} {\bibfnamefont {J.}~\bibnamefont {Zhu}}, \bibinfo {author}
  {\bibfnamefont {K.}~\bibnamefont {Watanabe}}, \bibinfo {author}
  {\bibfnamefont {T.}~\bibnamefont {Taniguchi}}, \bibinfo {author}
  {\bibfnamefont {L.}~\bibnamefont {Balents}}, \ and\ \bibinfo {author}
  {\bibfnamefont {A.~F.}\ \bibnamefont {Young}},\ }\bibfield  {title} {\enquote
  {\bibinfo {title} {Intrinsic quantized anomalous {H}all effect in a moir{\'e}
  heterostructure},}\ }\href {\doibase 10.1126/science.aay5533} {\bibfield
  {journal} {\bibinfo  {journal} {Science}\ }\textbf {\bibinfo {volume}
  {367}},\ \bibinfo {pages} {900} (\bibinfo {year} {2020})}\BibitemShut
  {NoStop}%
\bibitem [{\citenamefont {Chatterjee}\ \emph {et~al.}(2020)\citenamefont
  {Chatterjee}, \citenamefont {Bultinck},\ and\ \citenamefont
  {Zaletel}}]{Symmetry-breaking-2020}%
  \BibitemOpen
  \bibfield  {author} {\bibinfo {author} {\bibfnamefont {S.}~\bibnamefont
  {Chatterjee}}, \bibinfo {author} {\bibfnamefont {N.}~\bibnamefont
  {Bultinck}}, \ and\ \bibinfo {author} {\bibfnamefont {M.~P.}\ \bibnamefont
  {Zaletel}},\ }\bibfield  {title} {\enquote {\bibinfo {title} {{S}ymmetry
  breaking and skyrmionic transport in twisted bilayer graphene},}\ }\href
  {\doibase 10.1103/PhysRevB.101.165141} {\bibfield  {journal} {\bibinfo
  {journal} {Phys. Rev. B}\ }\textbf {\bibinfo {volume} {101}},\ \bibinfo
  {pages} {165141} (\bibinfo {year} {2020})}\BibitemShut {NoStop}%
\bibitem [{\citenamefont {Cea}\ \emph {et~al.}(2020)\citenamefont {Cea},
  \citenamefont {Pantale\'on},\ and\ \citenamefont
  {Guinea}}]{Band-structure-2020}%
  \BibitemOpen
  \bibfield  {author} {\bibinfo {author} {\bibfnamefont {T.}~\bibnamefont
  {Cea}}, \bibinfo {author} {\bibfnamefont {P.~A.}\ \bibnamefont
  {Pantale\'on}}, \ and\ \bibinfo {author} {\bibfnamefont {F.}~\bibnamefont
  {Guinea}},\ }\bibfield  {title} {\enquote {\bibinfo {title} {Band structure
  of twisted bilayer graphene on hexagonal boron nitride},}\ }\href {\doibase
  10.1103/PhysRevB.102.155136} {\bibfield  {journal} {\bibinfo  {journal}
  {Phys. Rev. B}\ }\textbf {\bibinfo {volume} {102}},\ \bibinfo {pages}
  {155136} (\bibinfo {year} {2020})}\BibitemShut {NoStop}%
\bibitem [{\citenamefont {Lin}\ \emph {et~al.}(2021{\natexlab{a}})\citenamefont
  {Lin}, \citenamefont {Su},\ and\ \citenamefont {Ni}}]{Misalignment2021Lin}%
  \BibitemOpen
  \bibfield  {author} {\bibinfo {author} {\bibfnamefont {X.}~\bibnamefont
  {Lin}}, \bibinfo {author} {\bibfnamefont {K.}~\bibnamefont {Su}}, \ and\
  \bibinfo {author} {\bibfnamefont {J.}~\bibnamefont {Ni}},\ }\bibfield
  {title} {\enquote {\bibinfo {title} {Misalignment instability in magic-angle
  twisted bilayer graphene on hexagonal boron nitride},}\ }\href {\doibase
  10.1088/2053-1583/abddcb} {\bibfield  {journal} {\bibinfo  {journal} {2D
  Mater.}\ }\textbf {\bibinfo {volume} {8}},\ \bibinfo {pages} {025025}
  (\bibinfo {year} {2021}{\natexlab{a}})}\BibitemShut {NoStop}%
\bibitem [{\citenamefont {Shi}\ \emph {et~al.}(2021)\citenamefont {Shi},
  \citenamefont {Zhu},\ and\ \citenamefont {MacDonald}}]{Moire2021Shi}%
  \BibitemOpen
  \bibfield  {author} {\bibinfo {author} {\bibfnamefont {J.}~\bibnamefont
  {Shi}}, \bibinfo {author} {\bibfnamefont {J.}~\bibnamefont {Zhu}}, \ and\
  \bibinfo {author} {\bibfnamefont {A.~H.}\ \bibnamefont {MacDonald}},\
  }\bibfield  {title} {\enquote {\bibinfo {title} {Moir\'e commensurability and
  the quantum anomalous {H}all effect in twisted bilayer graphene on hexagonal
  boron nitride},}\ }\href {\doibase 10.1103/PhysRevB.103.075122} {\bibfield
  {journal} {\bibinfo  {journal} {Phys. Rev. B}\ }\textbf {\bibinfo {volume}
  {103}},\ \bibinfo {pages} {075122} (\bibinfo {year} {2021})}\BibitemShut
  {NoStop}%
\bibitem [{\citenamefont {Mao}\ and\ \citenamefont
  {Senthil}(2021)}]{Quasiperiodicity2021Mao}%
  \BibitemOpen
  \bibfield  {author} {\bibinfo {author} {\bibfnamefont {D.}~\bibnamefont
  {Mao}}\ and\ \bibinfo {author} {\bibfnamefont {T.}~\bibnamefont {Senthil}},\
  }\bibfield  {title} {\enquote {\bibinfo {title} {Quasiperiodicity, band
  topology, and moir\'e graphene},}\ }\href {\doibase
  10.1103/PhysRevB.103.115110} {\bibfield  {journal} {\bibinfo  {journal}
  {Phys. Rev. B}\ }\textbf {\bibinfo {volume} {103}},\ \bibinfo {pages}
  {115110} (\bibinfo {year} {2021})}\BibitemShut {NoStop}%
\bibitem [{\citenamefont {Lopes~dos Santos}\ \emph {et~al.}(2007)\citenamefont
  {Lopes~dos Santos}, \citenamefont {Peres},\ and\ \citenamefont
  {Castro~Neto}}]{LopesdosSantos2007}%
  \BibitemOpen
  \bibfield  {author} {\bibinfo {author} {\bibfnamefont {J.~M.~B.}\
  \bibnamefont {Lopes~dos Santos}}, \bibinfo {author} {\bibfnamefont
  {N.~M.~R.}\ \bibnamefont {Peres}}, \ and\ \bibinfo {author} {\bibfnamefont
  {A.~H.}\ \bibnamefont {Castro~Neto}},\ }\bibfield  {title} {\enquote
  {\bibinfo {title} {Graphene {B}ilayer with a {T}wist: {E}lectronic
  {S}tructure},}\ }\href {\doibase 10.1103/PhysRevLett.99.256802} {\bibfield
  {journal} {\bibinfo  {journal} {Phys. Rev. Lett.}\ }\textbf {\bibinfo
  {volume} {99}},\ \bibinfo {pages} {256802} (\bibinfo {year}
  {2007})}\BibitemShut {NoStop}%
\bibitem [{\citenamefont {Mele}(2010)}]{Commensuration-and-2010}%
  \BibitemOpen
  \bibfield  {author} {\bibinfo {author} {\bibfnamefont {E.~J.}\ \bibnamefont
  {Mele}},\ }\bibfield  {title} {\enquote {\bibinfo {title} {{C}ommensuration
  and interlayer coherence in twisted bilayer graphene},}\ }\href {\doibase
  10.1103/PhysRevB.81.161405} {\bibfield  {journal} {\bibinfo  {journal} {Phys.
  Rev. B}\ }\textbf {\bibinfo {volume} {81}},\ \bibinfo {pages} {161405}
  (\bibinfo {year} {2010})}\BibitemShut {NoStop}%
\bibitem [{\citenamefont {Li}\ \emph {et~al.}(2010)\citenamefont {Li},
  \citenamefont {Luican}, \citenamefont {Lopes~dos Santos}, \citenamefont
  {Castro~Neto}, \citenamefont {Reina}, \citenamefont {Kong},\ and\
  \citenamefont {Andrei}}]{Observation-of-2010}%
  \BibitemOpen
  \bibfield  {author} {\bibinfo {author} {\bibfnamefont {G.}~\bibnamefont
  {Li}}, \bibinfo {author} {\bibfnamefont {A.}~\bibnamefont {Luican}}, \bibinfo
  {author} {\bibfnamefont {J.~M.~B.}\ \bibnamefont {Lopes~dos Santos}},
  \bibinfo {author} {\bibfnamefont {A.~H.}\ \bibnamefont {Castro~Neto}},
  \bibinfo {author} {\bibfnamefont {A.}~\bibnamefont {Reina}}, \bibinfo
  {author} {\bibfnamefont {J.}~\bibnamefont {Kong}}, \ and\ \bibinfo {author}
  {\bibfnamefont {E.~Y.}\ \bibnamefont {Andrei}},\ }\bibfield  {title}
  {\enquote {\bibinfo {title} {{O}bservation of {V}an {H}ove singularities in
  twisted graphene layers},}\ }\href {\doibase 10.1038/nphys1463} {\bibfield
  {journal} {\bibinfo  {journal} {Nat. Phys.}\ }\textbf {\bibinfo {volume}
  {6}},\ \bibinfo {pages} {109} (\bibinfo {year} {2010})}\BibitemShut {NoStop}%
\bibitem [{\citenamefont {Lopes~dos Santos}\ \emph {et~al.}(2012)\citenamefont
  {Lopes~dos Santos}, \citenamefont {Peres},\ and\ \citenamefont
  {Castro~Neto}}]{LopesdosSantos2012}%
  \BibitemOpen
  \bibfield  {author} {\bibinfo {author} {\bibfnamefont {J.~M.~B.}\
  \bibnamefont {Lopes~dos Santos}}, \bibinfo {author} {\bibfnamefont
  {N.~M.~R.}\ \bibnamefont {Peres}}, \ and\ \bibinfo {author} {\bibfnamefont
  {A.~H.}\ \bibnamefont {Castro~Neto}},\ }\bibfield  {title} {\enquote
  {\bibinfo {title} {Continuum model of the twisted graphene bilayer},}\ }\href
  {\doibase 10.1103/PhysRevB.86.155449} {\bibfield  {journal} {\bibinfo
  {journal} {Phys. Rev. B}\ }\textbf {\bibinfo {volume} {86}},\ \bibinfo
  {pages} {155449} (\bibinfo {year} {2012})}\BibitemShut {NoStop}%
\bibitem [{\citenamefont {Alden}\ \emph {et~al.}(2013)\citenamefont {Alden},
  \citenamefont {Tsen}, \citenamefont {Huang}, \citenamefont {Hovden},
  \citenamefont {Brown}, \citenamefont {Park}, \citenamefont {Muller},\ and\
  \citenamefont {McEuen}}]{McEuenBLG13}%
  \BibitemOpen
  \bibfield  {author} {\bibinfo {author} {\bibfnamefont {J.~S.}\ \bibnamefont
  {Alden}}, \bibinfo {author} {\bibfnamefont {A.~W.}\ \bibnamefont {Tsen}},
  \bibinfo {author} {\bibfnamefont {P.~Y.}\ \bibnamefont {Huang}}, \bibinfo
  {author} {\bibfnamefont {R.}~\bibnamefont {Hovden}}, \bibinfo {author}
  {\bibfnamefont {L.}~\bibnamefont {Brown}}, \bibinfo {author} {\bibfnamefont
  {J.}~\bibnamefont {Park}}, \bibinfo {author} {\bibfnamefont {D.~A.}\
  \bibnamefont {Muller}}, \ and\ \bibinfo {author} {\bibfnamefont {P.~L.}\
  \bibnamefont {McEuen}},\ }\bibfield  {title} {\enquote {\bibinfo {title}
  {Strain solitons and topological defects in bilayer graphene},}\ }\href
  {\doibase 10.1073/pnas.1309394110} {\bibfield  {journal} {\bibinfo  {journal}
  {Proc. Natl. Acad. Sci. U.S.A.}\ }\textbf {\bibinfo {volume} {110}},\
  \bibinfo {pages} {11256} (\bibinfo {year} {2013})}\BibitemShut {NoStop}%
\bibitem [{\citenamefont {Woods}\ \emph {et~al.}(2014)\citenamefont {Woods},
  \citenamefont {Britnell}, \citenamefont {Eckmann}, \citenamefont {Ma},
  \citenamefont {Lu}, \citenamefont {Guo}, \citenamefont {Lin}, \citenamefont
  {Yu}, \citenamefont {Cao}, \citenamefont {Gorbachev}, \citenamefont
  {Kretinin}, \citenamefont {Park}, \citenamefont {Ponomarenko}, \citenamefont
  {Katsnelson}, \citenamefont {Gornostyrev}, \citenamefont {Watanabe},
  \citenamefont {Taniguchi}, \citenamefont {Casiraghi}, \citenamefont {Gao},
  \citenamefont {Geim},\ and\ \citenamefont {Novoselov}}]{Commensurate2014}%
  \BibitemOpen
  \bibfield  {author} {\bibinfo {author} {\bibfnamefont {C.~R.}\ \bibnamefont
  {Woods}}, \bibinfo {author} {\bibfnamefont {L.}~\bibnamefont {Britnell}},
  \bibinfo {author} {\bibfnamefont {A.}~\bibnamefont {Eckmann}}, \bibinfo
  {author} {\bibfnamefont {R.~S.}\ \bibnamefont {Ma}}, \bibinfo {author}
  {\bibfnamefont {J.~C.}\ \bibnamefont {Lu}}, \bibinfo {author} {\bibfnamefont
  {H.~M.}\ \bibnamefont {Guo}}, \bibinfo {author} {\bibfnamefont
  {X.}~\bibnamefont {Lin}}, \bibinfo {author} {\bibfnamefont {G.~L.}\
  \bibnamefont {Yu}}, \bibinfo {author} {\bibfnamefont {Y.}~\bibnamefont
  {Cao}}, \bibinfo {author} {\bibfnamefont {R.~V.}\ \bibnamefont {Gorbachev}},
  \bibinfo {author} {\bibfnamefont {A.~V.}\ \bibnamefont {Kretinin}}, \bibinfo
  {author} {\bibfnamefont {J.}~\bibnamefont {Park}}, \bibinfo {author}
  {\bibfnamefont {L.~A.}\ \bibnamefont {Ponomarenko}}, \bibinfo {author}
  {\bibfnamefont {M.~I.}\ \bibnamefont {Katsnelson}}, \bibinfo {author}
  {\bibfnamefont {Y.~N.}\ \bibnamefont {Gornostyrev}}, \bibinfo {author}
  {\bibfnamefont {K.}~\bibnamefont {Watanabe}}, \bibinfo {author}
  {\bibfnamefont {T.}~\bibnamefont {Taniguchi}}, \bibinfo {author}
  {\bibfnamefont {C.}~\bibnamefont {Casiraghi}}, \bibinfo {author}
  {\bibfnamefont {H.-J.}\ \bibnamefont {Gao}}, \bibinfo {author} {\bibfnamefont
  {A.~K.}\ \bibnamefont {Geim}}, \ and\ \bibinfo {author} {\bibfnamefont
  {K.~S.}\ \bibnamefont {Novoselov}},\ }\bibfield  {title} {\enquote {\bibinfo
  {title} {{C}ommensurate{-}incommensurate transition in graphene on hexagonal
  boron nitride},}\ }\href {\doibase 10.1038/nphys2954} {\bibfield  {journal}
  {\bibinfo  {journal} {Nat. Phys.}\ }\textbf {\bibinfo {volume} {10}},\
  \bibinfo {pages} {451} (\bibinfo {year} {2014})}\BibitemShut {NoStop}%
\bibitem [{\citenamefont {Uchida}\ \emph {et~al.}(2014)\citenamefont {Uchida},
  \citenamefont {Furuya}, \citenamefont {Iwata},\ and\ \citenamefont
  {Oshiyama}}]{Uchida2014}%
  \BibitemOpen
  \bibfield  {author} {\bibinfo {author} {\bibfnamefont {K.}~\bibnamefont
  {Uchida}}, \bibinfo {author} {\bibfnamefont {S.}~\bibnamefont {Furuya}},
  \bibinfo {author} {\bibfnamefont {J.-I.}\ \bibnamefont {Iwata}}, \ and\
  \bibinfo {author} {\bibfnamefont {A.}~\bibnamefont {Oshiyama}},\ }\bibfield
  {title} {\enquote {\bibinfo {title} {Atomic corrugation and electron
  localization due to {M}oir{\'e} patterns in twisted bilayer graphenes},}\
  }\href {\doibase 10.1103/PhysRevB.90.155451} {\bibfield  {journal} {\bibinfo
  {journal} {Phys. Rev. B}\ }\textbf {\bibinfo {volume} {90}},\ \bibinfo
  {pages} {155451} (\bibinfo {year} {2014})}\BibitemShut {NoStop}%
\bibitem [{\citenamefont {San-Jose}\ \emph
  {et~al.}(2014{\natexlab{a}})\citenamefont {San-Jose}, \citenamefont
  {Guti\'{e}rrez-Rubio}, \citenamefont {Sturla},\ and\ \citenamefont
  {Guinea}}]{Electronic2014Sep}%
  \BibitemOpen
  \bibfield  {author} {\bibinfo {author} {\bibfnamefont {P.}~\bibnamefont
  {San-Jose}}, \bibinfo {author} {\bibfnamefont {A.}~\bibnamefont
  {Guti\'{e}rrez-Rubio}}, \bibinfo {author} {\bibfnamefont {M.}~\bibnamefont
  {Sturla}}, \ and\ \bibinfo {author} {\bibfnamefont {F.}~\bibnamefont
  {Guinea}},\ }\bibfield  {title} {\enquote {\bibinfo {title} {Electronic
  structure of spontaneously strained graphene on hexagonal boron nitride},}\
  }\href {\doibase 10.1103/PhysRevB.90.115152} {\bibfield  {journal} {\bibinfo
  {journal} {Phys. Rev. B}\ }\textbf {\bibinfo {volume} {90}},\ \bibinfo
  {pages} {115152} (\bibinfo {year} {2014}{\natexlab{a}})}\BibitemShut
  {NoStop}%
\bibitem [{\citenamefont {San-Jose}\ \emph
  {et~al.}(2014{\natexlab{b}})\citenamefont {San-Jose}, \citenamefont
  {Guti\'{e}rrez-Rubio}, \citenamefont {Sturla},\ and\ \citenamefont
  {Guinea}}]{Spontaneous2014Aug}%
  \BibitemOpen
  \bibfield  {author} {\bibinfo {author} {\bibfnamefont {P.}~\bibnamefont
  {San-Jose}}, \bibinfo {author} {\bibfnamefont {A.}~\bibnamefont
  {Guti\'{e}rrez-Rubio}}, \bibinfo {author} {\bibfnamefont {M.}~\bibnamefont
  {Sturla}}, \ and\ \bibinfo {author} {\bibfnamefont {F.}~\bibnamefont
  {Guinea}},\ }\bibfield  {title} {\enquote {\bibinfo {title} {Spontaneous
  strains and gap in graphene on boron nitride},}\ }\href {\doibase
  10.1103/PhysRevB.90.075428} {\bibfield  {journal} {\bibinfo  {journal} {Phys.
  Rev. B}\ }\textbf {\bibinfo {volume} {90}},\ \bibinfo {pages} {075428}
  (\bibinfo {year} {2014}{\natexlab{b}})}\BibitemShut {NoStop}%
\bibitem [{\citenamefont {Slotman}\ \emph {et~al.}(2015)\citenamefont
  {Slotman}, \citenamefont {van Wijk}, \citenamefont {Zhao}, \citenamefont
  {Fasolino}, \citenamefont {Katsnelson},\ and\ \citenamefont
  {Yuan}}]{Effect2015Oct}%
  \BibitemOpen
  \bibfield  {author} {\bibinfo {author} {\bibfnamefont {G.~J.}\ \bibnamefont
  {Slotman}}, \bibinfo {author} {\bibfnamefont {M.~M.}\ \bibnamefont {van
  Wijk}}, \bibinfo {author} {\bibfnamefont {P.-L.}\ \bibnamefont {Zhao}},
  \bibinfo {author} {\bibfnamefont {A.}~\bibnamefont {Fasolino}}, \bibinfo
  {author} {\bibfnamefont {M.~I.}\ \bibnamefont {Katsnelson}}, \ and\ \bibinfo
  {author} {\bibfnamefont {S.~J.}\ \bibnamefont {Yuan}},\ }\bibfield  {title}
  {\enquote {\bibinfo {title} {Effect of {S}tructural {R}elaxation on the
  {E}lectronic {S}tructure of {G}raphene on {H}exagonal {B}oron {N}itride},}\
  }\href {\doibase 10.1103/PhysRevLett.115.186801} {\bibfield  {journal}
  {\bibinfo  {journal} {Phys. Rev. Lett.}\ }\textbf {\bibinfo {volume} {115}},\
  \bibinfo {pages} {186801} (\bibinfo {year} {2015})}\BibitemShut {NoStop}%
\bibitem [{\citenamefont {Jung}\ \emph {et~al.}(2015)\citenamefont {Jung},
  \citenamefont {DaSilva}, \citenamefont {MacDonald},\ and\ \citenamefont
  {Adam}}]{Origin2015Feb}%
  \BibitemOpen
  \bibfield  {author} {\bibinfo {author} {\bibfnamefont {J.}~\bibnamefont
  {Jung}}, \bibinfo {author} {\bibfnamefont {A.~M.}\ \bibnamefont {DaSilva}},
  \bibinfo {author} {\bibfnamefont {A.~H.}\ \bibnamefont {MacDonald}}, \ and\
  \bibinfo {author} {\bibfnamefont {S.}~\bibnamefont {Adam}},\ }\bibfield
  {title} {\enquote {\bibinfo {title} {Origin of band gaps in graphene on
  hexagonal boron nitride},}\ }\href {\doibase 10.1038/ncomms7308} {\bibfield
  {journal} {\bibinfo  {journal} {Nat. Commun.}\ }\textbf {\bibinfo {volume}
  {6}},\ \bibinfo {pages} {6308} (\bibinfo {year} {2015})}\BibitemShut
  {NoStop}%
\bibitem [{\citenamefont {van Wijk}\ \emph {et~al.}(2015)\citenamefont {van
  Wijk}, \citenamefont {Schuring}, \citenamefont {Katsnelson},\ and\
  \citenamefont {Fasolino}}]{Wijk2015}%
  \BibitemOpen
  \bibfield  {author} {\bibinfo {author} {\bibfnamefont {M.~M.}\ \bibnamefont
  {van Wijk}}, \bibinfo {author} {\bibfnamefont {A.}~\bibnamefont {Schuring}},
  \bibinfo {author} {\bibfnamefont {M.~I.}\ \bibnamefont {Katsnelson}}, \ and\
  \bibinfo {author} {\bibfnamefont {A.}~\bibnamefont {Fasolino}},\ }\bibfield
  {title} {\enquote {\bibinfo {title} {Relaxation of {M}oir{\'e} patterns for
  slightly misaligned identical lattices: graphene on graphite},}\ }\href
  {\doibase 10.1088/2053-1583/2/3/034010} {\bibfield  {journal} {\bibinfo
  {journal} {2D Mater.}\ }\textbf {\bibinfo {volume} {2}},\ \bibinfo {pages}
  {034010} (\bibinfo {year} {2015})}\BibitemShut {NoStop}%
\bibitem [{\citenamefont {Jain}\ \emph {et~al.}(2017)\citenamefont {Jain},
  \citenamefont {Juri{\v{c}}i{\'c}},\ and\ \citenamefont {Barkema}}]{Jain2017}%
  \BibitemOpen
  \bibfield  {author} {\bibinfo {author} {\bibfnamefont {S.~K.}\ \bibnamefont
  {Jain}}, \bibinfo {author} {\bibfnamefont {V.}~\bibnamefont
  {Juri{\v{c}}i{\'c}}}, \ and\ \bibinfo {author} {\bibfnamefont {G.~T.}\
  \bibnamefont {Barkema}},\ }\bibfield  {title} {\enquote {\bibinfo {title}
  {Structure of twisted and buckled bilayer graphene},}\ }\href {\doibase
  10.1088/2053-1583/4/1/015018} {\bibfield  {journal} {\bibinfo  {journal} {2D
  Mater.}\ }\textbf {\bibinfo {volume} {4}},\ \bibinfo {pages} {015018}
  (\bibinfo {year} {2017})}\BibitemShut {NoStop}%
\bibitem [{\citenamefont {Jung}\ \emph {et~al.}(2017)\citenamefont {Jung},
  \citenamefont {Laksono}, \citenamefont {DaSilva}, \citenamefont {MacDonald},
  \citenamefont {Mucha-Kruczy\'{n}ski},\ and\ \citenamefont
  {Adam}}]{Moire2017Aug}%
  \BibitemOpen
  \bibfield  {author} {\bibinfo {author} {\bibfnamefont {J.}~\bibnamefont
  {Jung}}, \bibinfo {author} {\bibfnamefont {E.}~\bibnamefont {Laksono}},
  \bibinfo {author} {\bibfnamefont {A.~M.}\ \bibnamefont {DaSilva}}, \bibinfo
  {author} {\bibfnamefont {A.~H.}\ \bibnamefont {MacDonald}}, \bibinfo {author}
  {\bibfnamefont {M.}~\bibnamefont {Mucha-Kruczy\'{n}ski}}, \ and\ \bibinfo
  {author} {\bibfnamefont {S.}~\bibnamefont {Adam}},\ }\bibfield  {title}
  {\enquote {\bibinfo {title} {Moir\'{e} band model and band gaps of graphene
  on hexagonal boron nitride},}\ }\href {\doibase 10.1103/PhysRevB.96.085442}
  {\bibfield  {journal} {\bibinfo  {journal} {Phys. Rev. B}\ }\textbf {\bibinfo
  {volume} {96}},\ \bibinfo {pages} {085442} (\bibinfo {year}
  {2017})}\BibitemShut {NoStop}%
\bibitem [{\citenamefont {Gargiulo}\ and\ \citenamefont
  {Yazyev}(2018)}]{Gargiulo2018}%
  \BibitemOpen
  \bibfield  {author} {\bibinfo {author} {\bibfnamefont {F.}~\bibnamefont
  {Gargiulo}}\ and\ \bibinfo {author} {\bibfnamefont {O.~V.}\ \bibnamefont
  {Yazyev}},\ }\bibfield  {title} {\enquote {\bibinfo {title} {Structural and
  electronic transformation in low-angle twisted bilayer graphene},}\ }\href
  {\doibase 10.1088/2053-1583/aa9640} {\bibfield  {journal} {\bibinfo
  {journal} {2D Mater.}\ }\textbf {\bibinfo {volume} {5}},\ \bibinfo {pages}
  {015019} (\bibinfo {year} {2018})}\BibitemShut {NoStop}%
\bibitem [{\citenamefont {Carr}\ \emph {et~al.}(2018)\citenamefont {Carr},
  \citenamefont {Massatt}, \citenamefont {Torrisi}, \citenamefont {Cazeaux},
  \citenamefont {Luskin},\ and\ \citenamefont {Kaxiras}}]{carr2018relaxation}%
  \BibitemOpen
  \bibfield  {author} {\bibinfo {author} {\bibfnamefont {S.}~\bibnamefont
  {Carr}}, \bibinfo {author} {\bibfnamefont {D.}~\bibnamefont {Massatt}},
  \bibinfo {author} {\bibfnamefont {S.~B.}\ \bibnamefont {Torrisi}}, \bibinfo
  {author} {\bibfnamefont {P.}~\bibnamefont {Cazeaux}}, \bibinfo {author}
  {\bibfnamefont {M.}~\bibnamefont {Luskin}}, \ and\ \bibinfo {author}
  {\bibfnamefont {E.}~\bibnamefont {Kaxiras}},\ }\bibfield  {title} {\enquote
  {\bibinfo {title} {Relaxation and domain formation in incommensurate
  two-dimensional heterostructures},}\ }\href {\doibase
  10.1103/PhysRevB.98.224102} {\bibfield  {journal} {\bibinfo  {journal} {Phys.
  Rev. B}\ }\textbf {\bibinfo {volume} {98}},\ \bibinfo {pages} {224102}
  (\bibinfo {year} {2018})}\BibitemShut {NoStop}%
\bibitem [{\citenamefont {Qiao}\ \emph {et~al.}(2018)\citenamefont {Qiao},
  \citenamefont {Yin},\ and\ \citenamefont {He}}]{Twisted-graphene-2018}%
  \BibitemOpen
  \bibfield  {author} {\bibinfo {author} {\bibfnamefont {J.-B.}\ \bibnamefont
  {Qiao}}, \bibinfo {author} {\bibfnamefont {L.-J.}\ \bibnamefont {Yin}}, \
  and\ \bibinfo {author} {\bibfnamefont {L.}~\bibnamefont {He}},\ }\bibfield
  {title} {\enquote {\bibinfo {title} {{T}wisted graphene bilayer around the
  first magic angle engineered by heterostrain},}\ }\href {\doibase
  10.1103/PhysRevB.98.235402} {\bibfield  {journal} {\bibinfo  {journal} {Phys.
  Rev. B}\ }\textbf {\bibinfo {volume} {98}},\ \bibinfo {pages} {235402}
  (\bibinfo {year} {2018})}\BibitemShut {NoStop}%
\bibitem [{\citenamefont {Lin}\ and\ \citenamefont
  {Ni}(2019)}]{Effective2019lin}%
  \BibitemOpen
  \bibfield  {author} {\bibinfo {author} {\bibfnamefont {X.}~\bibnamefont
  {Lin}}\ and\ \bibinfo {author} {\bibfnamefont {J.}~\bibnamefont {Ni}},\
  }\bibfield  {title} {\enquote {\bibinfo {title} {Effective lattice model of
  graphene moir\'e superlattices on hexagonal boron nitride},}\ }\href
  {\doibase 10.1103/PhysRevB.100.195413} {\bibfield  {journal} {\bibinfo
  {journal} {Phys. Rev. B}\ }\textbf {\bibinfo {volume} {100}},\ \bibinfo
  {pages} {195413} (\bibinfo {year} {2019})}\BibitemShut {NoStop}%
\bibitem [{\citenamefont {Liu}\ \emph {et~al.}(2020)\citenamefont {Liu},
  \citenamefont {Su}, \citenamefont {Zhou}, \citenamefont {Yin}, \citenamefont
  {Yan}, \citenamefont {Li}, \citenamefont {Yan}, \citenamefont {Han},
  \citenamefont {Fu}, \citenamefont {Zhang}, \citenamefont {Yang},
  \citenamefont {Ren},\ and\ \citenamefont {He}}]{Tunable-Lattice-2020}%
  \BibitemOpen
  \bibfield  {author} {\bibinfo {author} {\bibfnamefont {Y.-W.}\ \bibnamefont
  {Liu}}, \bibinfo {author} {\bibfnamefont {Y.}~\bibnamefont {Su}}, \bibinfo
  {author} {\bibfnamefont {X.-F.}\ \bibnamefont {Zhou}}, \bibinfo {author}
  {\bibfnamefont {L.-J.}\ \bibnamefont {Yin}}, \bibinfo {author} {\bibfnamefont
  {C.}~\bibnamefont {Yan}}, \bibinfo {author} {\bibfnamefont {S.-Y.}\
  \bibnamefont {Li}}, \bibinfo {author} {\bibfnamefont {W.}~\bibnamefont
  {Yan}}, \bibinfo {author} {\bibfnamefont {S.}~\bibnamefont {Han}}, \bibinfo
  {author} {\bibfnamefont {Z.-Q.}\ \bibnamefont {Fu}}, \bibinfo {author}
  {\bibfnamefont {Y.}~\bibnamefont {Zhang}}, \bibinfo {author} {\bibfnamefont
  {Q.}~\bibnamefont {Yang}}, \bibinfo {author} {\bibfnamefont {Y.-N.}\
  \bibnamefont {Ren}}, \ and\ \bibinfo {author} {\bibfnamefont
  {L.}~\bibnamefont {He}},\ }\bibfield  {title} {\enquote {\bibinfo {title}
  {{T}unable {L}attice {R}econstruction, {T}riangular {N}etwork of {C}hiral
  {O}ne-{D}imensional {S}tates, and {B}andwidth of {F}lat {B}ands in {M}agic
  {A}ngle {T}wisted {B}ilayer {G}raphene},}\ }\href {\doibase
  10.1103/PhysRevLett.125.236102} {\bibfield  {journal} {\bibinfo  {journal}
  {Phys. Rev. Lett.}\ }\textbf {\bibinfo {volume} {125}},\ \bibinfo {pages}
  {236102} (\bibinfo {year} {2020})}\BibitemShut {NoStop}%
\bibitem [{\citenamefont {Halbertal}\ \emph {et~al.}(2021)\citenamefont
  {Halbertal}, \citenamefont {Finney}, \citenamefont {Sunku}, \citenamefont
  {Kerelsky}, \citenamefont {Rubio-Verd\'{u}}, \citenamefont {Shabani},
  \citenamefont {Xian}, \citenamefont {Carr}, \citenamefont {Chen},
  \citenamefont {Zhang}, \citenamefont {Wang}, \citenamefont
  {Gonzalez-Acevedo}, \citenamefont {McLeod}, \citenamefont {Rhodes},
  \citenamefont {Watanabe}, \citenamefont {Taniguchi}, \citenamefont {Kaxiras},
  \citenamefont {Dean}, \citenamefont {Hone}, \citenamefont {Pasupathy},
  \citenamefont {Kennes}, \citenamefont {Rubio},\ and\ \citenamefont
  {Basov}}]{Moire-metrology-2021}%
  \BibitemOpen
  \bibfield  {author} {\bibinfo {author} {\bibfnamefont {D.}~\bibnamefont
  {Halbertal}}, \bibinfo {author} {\bibfnamefont {N.~R.}\ \bibnamefont
  {Finney}}, \bibinfo {author} {\bibfnamefont {S.~S.}\ \bibnamefont {Sunku}},
  \bibinfo {author} {\bibfnamefont {A.}~\bibnamefont {Kerelsky}}, \bibinfo
  {author} {\bibfnamefont {C.}~\bibnamefont {Rubio-Verd\'{u}}}, \bibinfo
  {author} {\bibfnamefont {S.}~\bibnamefont {Shabani}}, \bibinfo {author}
  {\bibfnamefont {L.}~\bibnamefont {Xian}}, \bibinfo {author} {\bibfnamefont
  {S.}~\bibnamefont {Carr}}, \bibinfo {author} {\bibfnamefont {S.}~\bibnamefont
  {Chen}}, \bibinfo {author} {\bibfnamefont {C.}~\bibnamefont {Zhang}},
  \bibinfo {author} {\bibfnamefont {L.}~\bibnamefont {Wang}}, \bibinfo {author}
  {\bibfnamefont {D.}~\bibnamefont {Gonzalez-Acevedo}}, \bibinfo {author}
  {\bibfnamefont {A.~S.}\ \bibnamefont {McLeod}}, \bibinfo {author}
  {\bibfnamefont {D.}~\bibnamefont {Rhodes}}, \bibinfo {author} {\bibfnamefont
  {K.}~\bibnamefont {Watanabe}}, \bibinfo {author} {\bibfnamefont
  {T.}~\bibnamefont {Taniguchi}}, \bibinfo {author} {\bibfnamefont
  {E.}~\bibnamefont {Kaxiras}}, \bibinfo {author} {\bibfnamefont {C.~R.}\
  \bibnamefont {Dean}}, \bibinfo {author} {\bibfnamefont {J.~C.}\ \bibnamefont
  {Hone}}, \bibinfo {author} {\bibfnamefont {A.~N.}\ \bibnamefont {Pasupathy}},
  \bibinfo {author} {\bibfnamefont {D.~M.}\ \bibnamefont {Kennes}}, \bibinfo
  {author} {\bibfnamefont {A.}~\bibnamefont {Rubio}}, \ and\ \bibinfo {author}
  {\bibfnamefont {D.~N.}\ \bibnamefont {Basov}},\ }\bibfield  {title} {\enquote
  {\bibinfo {title} {{M}oir\'{e} metrology of energy landscapes in van der
  {W}aals heterostructures},}\ }\href {\doibase 10.1038/s41467-020-20428-1}
  {\bibfield  {journal} {\bibinfo  {journal} {Nat. Commun.}\ }\textbf {\bibinfo
  {volume} {12}},\ \bibinfo {pages} {242} (\bibinfo {year} {2021})}\BibitemShut
  {NoStop}%
\bibitem [{\citenamefont {Gadelha}\ \emph {et~al.}(2021)\citenamefont
  {Gadelha}, \citenamefont {Ohlberg}, \citenamefont {Rabelo}, \citenamefont
  {Neto}, \citenamefont {Vasconcelos}, \citenamefont {Campos}, \citenamefont
  {Lemos}, \citenamefont {Ornelas}, \citenamefont {Miranda}, \citenamefont
  {Nadas}, \citenamefont {Santana}, \citenamefont {Watanabe}, \citenamefont
  {Taniguchi}, \citenamefont {van Troeye}, \citenamefont {Lamparski},
  \citenamefont {Meunier}, \citenamefont {Nguyen}, \citenamefont {Paszko},
  \citenamefont {Charlier}, \citenamefont {Campos}, \citenamefont {Cançado},
  \citenamefont {Medeiros-Ribeiro},\ and\ \citenamefont
  {Jorio}}]{Localization-of-2021}%
  \BibitemOpen
  \bibfield  {author} {\bibinfo {author} {\bibfnamefont {A.~C.}\ \bibnamefont
  {Gadelha}}, \bibinfo {author} {\bibfnamefont {D.~A.~A.}\ \bibnamefont
  {Ohlberg}}, \bibinfo {author} {\bibfnamefont {C.}~\bibnamefont {Rabelo}},
  \bibinfo {author} {\bibfnamefont {E.~G.~S.}\ \bibnamefont {Neto}}, \bibinfo
  {author} {\bibfnamefont {T.~L.}\ \bibnamefont {Vasconcelos}}, \bibinfo
  {author} {\bibfnamefont {J.~L.}\ \bibnamefont {Campos}}, \bibinfo {author}
  {\bibfnamefont {J.~S.}\ \bibnamefont {Lemos}}, \bibinfo {author}
  {\bibfnamefont {V.}~\bibnamefont {Ornelas}}, \bibinfo {author} {\bibfnamefont
  {D.}~\bibnamefont {Miranda}}, \bibinfo {author} {\bibfnamefont
  {R.}~\bibnamefont {Nadas}}, \bibinfo {author} {\bibfnamefont {F.~C.}\
  \bibnamefont {Santana}}, \bibinfo {author} {\bibfnamefont {K.}~\bibnamefont
  {Watanabe}}, \bibinfo {author} {\bibfnamefont {T.}~\bibnamefont {Taniguchi}},
  \bibinfo {author} {\bibfnamefont {B.}~\bibnamefont {van Troeye}}, \bibinfo
  {author} {\bibfnamefont {M.}~\bibnamefont {Lamparski}}, \bibinfo {author}
  {\bibfnamefont {V.}~\bibnamefont {Meunier}}, \bibinfo {author} {\bibfnamefont
  {V.-H.}\ \bibnamefont {Nguyen}}, \bibinfo {author} {\bibfnamefont
  {D.}~\bibnamefont {Paszko}}, \bibinfo {author} {\bibfnamefont {J.-C.}\
  \bibnamefont {Charlier}}, \bibinfo {author} {\bibfnamefont {L.~C.}\
  \bibnamefont {Campos}}, \bibinfo {author} {\bibfnamefont {L.~G.}\
  \bibnamefont {Cançado}}, \bibinfo {author} {\bibfnamefont {G.}~\bibnamefont
  {Medeiros-Ribeiro}}, \ and\ \bibinfo {author} {\bibfnamefont
  {A.}~\bibnamefont {Jorio}},\ }\bibfield  {title} {\enquote {\bibinfo {title}
  {{L}ocalization of lattice dynamics in low-angle twisted bilayer graphene},}\
  }\href {\doibase 10.1038/s41586-021-03252-5} {\bibfield  {journal} {\bibinfo
  {journal} {Nature}\ }\textbf {\bibinfo {volume} {590}},\ \bibinfo {pages}
  {405} (\bibinfo {year} {2021})}\BibitemShut {NoStop}%
\bibitem [{\citenamefont {Lin}\ \emph {et~al.}(2021{\natexlab{b}})\citenamefont
  {Lin}, \citenamefont {Zhu},\ and\ \citenamefont {Ni}}]{Emergence-of-2021}%
  \BibitemOpen
  \bibfield  {author} {\bibinfo {author} {\bibfnamefont {X.}~\bibnamefont
  {Lin}}, \bibinfo {author} {\bibfnamefont {H.}~\bibnamefont {Zhu}}, \ and\
  \bibinfo {author} {\bibfnamefont {J.}~\bibnamefont {Ni}},\ }\bibfield
  {title} {\enquote {\bibinfo {title} {{E}mergence of intrinsically isolated
  flat bands and their topology in fully relaxed twisted multilayer
  graphene},}\ }\href {\doibase 10.1103/PhysRevB.104.125421} {\bibfield
  {journal} {\bibinfo  {journal} {Phys. Rev. B}\ }\textbf {\bibinfo {volume}
  {104}},\ \bibinfo {pages} {125421} (\bibinfo {year}
  {2021}{\natexlab{b}})}\BibitemShut {NoStop}%
\bibitem [{\citenamefont {McGilly}\ \emph {et~al.}(2020)\citenamefont
  {McGilly}, \citenamefont {Kerelsky}, \citenamefont {Finney}, \citenamefont
  {Shapovalov}, \citenamefont {Shih}, \citenamefont {Ghiotto}, \citenamefont
  {Zeng}, \citenamefont {Moore}, \citenamefont {Wu}, \citenamefont {Bai},
  \citenamefont {Watanabe}, \citenamefont {Taniguchi}, \citenamefont {Stengel},
  \citenamefont {Zhou}, \citenamefont {Hone}, \citenamefont {Zhu},
  \citenamefont {Basov}, \citenamefont {Dean}, \citenamefont {Dreyer},\ and\
  \citenamefont {Pasupathy}}]{Visualization-of-2020}%
  \BibitemOpen
  \bibfield  {author} {\bibinfo {author} {\bibfnamefont {L.~J.}\ \bibnamefont
  {McGilly}}, \bibinfo {author} {\bibfnamefont {A.}~\bibnamefont {Kerelsky}},
  \bibinfo {author} {\bibfnamefont {N.~R.}\ \bibnamefont {Finney}}, \bibinfo
  {author} {\bibfnamefont {K.}~\bibnamefont {Shapovalov}}, \bibinfo {author}
  {\bibfnamefont {E.-M.}\ \bibnamefont {Shih}}, \bibinfo {author}
  {\bibfnamefont {A.}~\bibnamefont {Ghiotto}}, \bibinfo {author} {\bibfnamefont
  {Y.}~\bibnamefont {Zeng}}, \bibinfo {author} {\bibfnamefont {S.~L.}\
  \bibnamefont {Moore}}, \bibinfo {author} {\bibfnamefont {W.}~\bibnamefont
  {Wu}}, \bibinfo {author} {\bibfnamefont {Y.}~\bibnamefont {Bai}}, \bibinfo
  {author} {\bibfnamefont {K.}~\bibnamefont {Watanabe}}, \bibinfo {author}
  {\bibfnamefont {T.}~\bibnamefont {Taniguchi}}, \bibinfo {author}
  {\bibfnamefont {M.}~\bibnamefont {Stengel}}, \bibinfo {author} {\bibfnamefont
  {L.}~\bibnamefont {Zhou}}, \bibinfo {author} {\bibfnamefont {J.}~\bibnamefont
  {Hone}}, \bibinfo {author} {\bibfnamefont {X.}~\bibnamefont {Zhu}}, \bibinfo
  {author} {\bibfnamefont {D.~N.}\ \bibnamefont {Basov}}, \bibinfo {author}
  {\bibfnamefont {C.}~\bibnamefont {Dean}}, \bibinfo {author} {\bibfnamefont
  {C.~E.}\ \bibnamefont {Dreyer}}, \ and\ \bibinfo {author} {\bibfnamefont
  {A.~N.}\ \bibnamefont {Pasupathy}},\ }\bibfield  {title} {\enquote {\bibinfo
  {title} {{V}isualization of moiré superlattices},}\ }\href {\doibase
  10.1038/s41565-020-0708-3} {\bibfield  {journal} {\bibinfo  {journal} {Nat.
  Nanotechnol.}\ }\textbf {\bibinfo {volume} {15}},\ \bibinfo {pages} {580}
  (\bibinfo {year} {2020})}\BibitemShut {NoStop}%
\bibitem [{\citenamefont {Li}\ \emph {et~al.}()\citenamefont {Li},
  \citenamefont {Wang}, \citenamefont {Tang}, \citenamefont {Wang},
  \citenamefont {Watanabe}, \citenamefont {Taniguchi}, \citenamefont
  {R.~Gamelin}, \citenamefont {H.~Cobden}, \citenamefont {Yankowitz},
  \citenamefont {Xu},\ and\ \citenamefont {Li}}]{Unraveling-intrinsic-2021}%
  \BibitemOpen
  \bibfield  {author} {\bibinfo {author} {\bibfnamefont {Y.}~\bibnamefont
  {Li}}, \bibinfo {author} {\bibfnamefont {X.}~\bibnamefont {Wang}}, \bibinfo
  {author} {\bibfnamefont {D.}~\bibnamefont {Tang}}, \bibinfo {author}
  {\bibfnamefont {X.}~\bibnamefont {Wang}}, \bibinfo {author} {\bibfnamefont
  {K.}~\bibnamefont {Watanabe}}, \bibinfo {author} {\bibfnamefont
  {T.}~\bibnamefont {Taniguchi}}, \bibinfo {author} {\bibfnamefont
  {D.}~\bibnamefont {R.~Gamelin}}, \bibinfo {author} {\bibfnamefont
  {D.}~\bibnamefont {H.~Cobden}}, \bibinfo {author} {\bibfnamefont
  {M.}~\bibnamefont {Yankowitz}}, \bibinfo {author} {\bibfnamefont
  {X.}~\bibnamefont {Xu}}, \ and\ \bibinfo {author} {\bibfnamefont
  {J.}~\bibnamefont {Li}},\ }\bibfield  {title} {\enquote {\bibinfo {title}
  {{U}nraveling intrinsic flexoelectricity in twisted double bilayer
  graphene},}\ }\href {https://arxiv.org/abs/2104.02401} {\bibinfo  {journal}
  {arXiv:2104.02401}\ }\BibitemShut {NoStop}%
\end{thebibliography}

%

\end{document}